\documentclass[fleqn,usenatbib]{mnras}
\usepackage[T1]{fontenc}
\usepackage{newtxtext,newtxmath}
\DeclareRobustCommand{\VAN}[3]{#2}
\let\VANthebibliography\thebibliography
\def\thebibliography{\DeclareRobustCommand{\VAN}[3]{##3}\VANthebibliography}

\usepackage{graphicx}	% Including figure files
\usepackage{amsmath}	% Advanced maths commands
\usepackage{lineno}
\usepackage{multirow}
\usepackage{subcaption}
\usepackage{pdflscape}
\usepackage{orcidlink}
\title[The NGC 205 Variable Star Catalogue]{Variable Stars in Local Group Galaxies IX: Tracing the Complex History of NGC 205}

\author[I. Plaks et al.]{
Irina Plaks,$^{1}$\thanks{E-mail: iplaks@iastate.edu}\orcidlink{0009-0003-3216-3402}
M. Marengo,$^{2}$\orcidlink{0000-0001-9910-9230}
M. Monelli,$^{3,4,5}$\orcidlink{0000-0001-5292-6380}
C. R. Kerton,$^{1}$\orcidlink{0000-0003-1539-3321}
Giuliana Fiorentino,$^{6}$\orcidlink{0000-0003-0376-6928}
G. Bono,$^{7}$\orcidlink{0000-0001-5292-6380}
\newauthor
Clara E. Mart\'inez-V\'azquez,$^{8}$\thanks{E-mail: clara.martinez@noirlab.edu}\orcidlink{0000-0002-9144-7726}
\\
$^{1}$Department of Physics and Astronomy, Iowa State University, Ames, IA 50011, USA\\
$^{2}$Florida State University, Department of Physics, 77 Chieftain Way, Tallahassee, FL 32306, USA\\
$^{3}$INAF - Osservatorio Astronomico d'Abruzzo, via Mentore Maggini s.n.c., 64100 Teramo, Italy\\
$^{4}$IAC - Instituto de Astrofísica de Canarias, Calle Vía Lactea s/n, E-38205 La Laguna, Tenerife, Spain\\
$^{5}$Departmento de Astrofísica, Universidad de La Laguna, E-38206 La Laguna, Tenerife, Spain\\
$^{6}$INAF-Osservatorio Astronomico di Roma, via Frascati 33, 00078 Monte Porzio Catone, Italy\\
$^{7}$Department of Physics, Università di Roma Tor Vergata, Via della Ricerca Scientifica 1, 00133 Roma, Italy\\
$^{8}$NSF NOIRLab, 670 N. A'ohoku Place, Hilo, HI 96720, USA\\
}

\date{Accepted XXX. Received YYY; in original form ZZZ}

\pubyear{\the\year{}}

\begin{document}
\label{firstpage}
\pagerange{\pageref{firstpage}--\pageref{lastpage}}
\maketitle

\begin{abstract}
We present the most extensive catalogue of variable stars in NGC 205 to date, based on archival HST/ACS imaging of a single field ${\sim} 4'$ from the galaxy centre (GO-15902). We detect 1441 RR Lyrae stars (1064 RRab, 363 RRc, and 14 of uncertain mode) and 35 anomalous Cepheids, and measure their periods, amplitudes, and intensity-mean magnitudes in the F606W and F814W bands. The variable population is dominated by RR Lyrae stars, as expected for a predominantly old ($>$10 Gyr) field with no strong young component. From the red giant branch tip we derive a distance modulus of \(\mu = 24.68 \pm 0.06\) mag. The distribution of the RR Lyrae stars in the Bailey diagram, the fraction of RRc stars, and the presence of a substantial population of high-amplitude short-period RRab stars, indicate a range of metallicities extending to a metal-rich component, while the ratio of RR Lyrae stars to red giant branch stars indicates an extended star formation history, with only a modest fraction of the stellar mass formed in the first few Gyr. The anomalous Cepheids place NGC 205 among the slow evolvers on the relation between their frequency and host luminosity, and comparison with evolutionary models constrains their progenitors to $[\mathrm{Fe/H}] \lesssim -2.1$, indicating that metal-poor gas remained available well after the galaxy had begun to enrich. In a forthcoming paper, we will use the RR Lyrae periods and intensity-mean magnitudes to derive individual metallicities for these stars and trace the chemical enrichment history of NGC 205.

\end{abstract}

\begin{keywords}
stars: variables: RR Lyrae -- stars: variables: Cepheids -- galaxies: dwarf -- galaxies: stellar content -- galaxies: star formation -- Local Group
\end{keywords}

%%%%%%%%%%%%%%%%%%%%%%%%%%%%%%%%%%%%%%%%%%%%%%%%%%

%%%%%%%%%%%%%%%%% BODY OF PAPER %%%%%%%%%%%%%%%%%%

%\linenumbers
\section{Introduction} \label{sec:intro}

Understanding the formation and evolution of early-type galaxies (ETG) is essential for constraining galaxy assembly in the early Universe. As some of the oldest stellar systems, they hold important clues about early star formation and chemical evolution. This is especially true for dwarf ellipticals (dE), the most common population in galaxy clusters, as they are thought to be the building blocks of more massive galaxies \citep{White-1978,White-1991}. Previous studies have shown that ETGs host stars with different ages, metallicities, and kinematic properties (e.g. \citealt{Kuntschner-2010,Thomas-2010,Cappellari-2013b,Lu-2023}), indicating complex evolutionary histories. These observations are consistent with a two-phase formation scenario, in agreement with Lambda Cold Dark Matter ($\Lambda$CDM), wherein a very early dissipative phase forms a galactic core and is followed by a second phase driven by mergers and accretion in which it grows in mass and radius \citep{oser-2010}. At the same time, other fundamental questions concerning the stellar population of ETGs remain unresolved, including whether their initial mass function is significantly different from the one inferred in the Milky Way \citep{cappellari-2012}.

NGC 205, the closest and largest dE satellite of M31, is particularly well suited to investigate these processes. It has very likely experienced tidal interaction and stripping from M31 \citep{Choi-2022}, and hosts both an ancient stellar population as well as young blue stars near its center, demonstrating precisely the complex and extended star formation history predicted by the two-phase scenario. Furthermore, its proximity allows individual stars to be resolved, enabling direct measurements of stellar properties such as age and metallicity. NGC 205 is therefore key to understanding the origin and evolution of ETGs, and provides valuable insight into how dEs assemble and evolve as satellites of massive host galaxies.

Carrying out formation and evolution studies using a galaxy’s resolved stellar population first requires choosing an appropriate tracer. The choice of tracer is essential as stars must be observed photometrically with relative ease and must also reflect the stellar populations being studied. RR Lyrae (RRL) stars are one such tracer. They are a class of pulsating variables, meaning that their magnitude changes with time according to a regular period (typically between $0.2$ and $1.0$ d). RRLs are low-mass ($0.6 - 0.8$ \(\mathrm{M}_\odot\)) evolved stars that lie in the instability strip on the horizontal branch (HB) and are at least 10 Gyr old. 

Given the extreme age of these stars, along with their bright and narrowly defined absolute magnitude ($M_{V} \approx 0.6$ mag) and distinctive light curves, RRLs are excellent probes of a galaxy’s oldest stellar populations \citep{layden-1995}. Their magnitude and distinctive light curves make them easily identifiable, and their position on the HB makes them easily distinguishable from other variable stars. Furthermore, the existence of RRL period-luminosity-metallicity (PLZ) and period-Wesenheit-metallicity (PWZ) relations make it possible to accurately determine their distances and metallicities when other frequently used methods are prohibitive.

%The characteristic pulsation of RRL stars is believed to be caused mainly by the $\kappa$ mechanism, wherein the opacity ($\kappa$) of a star’s surface layers is dependent on temperature and density. These pulsations are restricted to the outer envelope and have essentially no effect upon the stellar interior. In the case of RRL stars, this pulsation occurs in the stars’ helium ionization zone. Within this region, gas becomes most opaque near maximum compression (and maximum temperature) due to ionization of the helium, trapping radiation and resulting in a build-up of pressure that causes the star's surface to expand \citep{Smith-1995}. Expansion causes a decrease in density and temperature, and therefore in opacity, allowing the trapped radiation to escape and pressure to decrease. Gravity then compresses the star back to its original size. This cycle repeats, leading to regular changes in the star’s magnitude.

The characteristic pulsation of RRL stars can occur in different modes, which categorises them into three main types. Fundamental mode pulsators (RRab) are the most common type, and can be recognised by their asymmetric light curve, caused by a rapid increase in light followed by a much slower decrease, and longer periods (typically, 0.45 -- 1.0 d). The next most common are first-overtone pulsators (RRc), which have one pulsation node in their interior. RRc stars, unlike RRab stars, have an equally or nearly equally rapid increase and decrease in light, and thus have much more symmetric, sinusoidal light curves, as well as shorter periods (typically, 0.2 -- 0.45 d). There are also a small number of RRL stars which pulsate in both the fundamental and first-overtone modes, known as RRd stars, which are thought to be in the process of switching between the other two classes. 

Anomalous Cepheids (ACs) complement RRLs as tracers of a galaxy’s history. Where RRLs probe the ancient population alone, ACs can constrain the age and metallicity of intermediate-age stars (though some are formed as the result of merging old stars). Like RRLs, they are pulsators that cross the instability strip while burning helium in their cores, but are more massive at $1-2 M_{\odot}$ and ignite helium under conditions of partial electron degeneracy, requiring a very low metallicity ($Z \lesssim 0.0008$) for their evolutionary tracks to reach the strip \citep{Renzini-1977, Fiorentino-2012}. Consequently, ACs trace the metal-poor gas available at the epoch of their formation. As their higher mass suggests, they are brighter than RRLs, sitting at ${\sim}1$ mag above the HB, and are considerably rarer. They pulsate in either the fundamental or first-overtone mode, but their light curve shapes do not map onto the mode as cleanly as those of RRLs. Instead, mode discrimination requires period, colour, and amplitude together \citep{Marconi-2004}.

Variable stars in NGC 205 have been identified in several previous studies (e.g. \citealt{savino-2022,sarajedini-2025}), establishing the presence of a substantial RRL population. However, the published catalogues are incomplete, limiting the statistical power of subsequent analyses. In this paper, we present a substantially expanded catalogue containing over 1400 RRL stars, along with a small population of anomalous Cepheids and several unclassified variables, more than tripling the number of known variables in NGC 205. This sample provides a more complete census of the galaxy's ancient stellar population and establishes the foundation for improved measurements of its chemical enrichment history and star formation history, which will be presented in a future paper in this series.

This paper focuses on the identification and characterisation of the variable star population in NGC 205. We describe the methods used to find and classify variable stars and report their fundamental properties, including pulsation period, mode, amplitude, and intensity-mean magnitude. The paper is organised as follows. Section~\ref{sec:data} describes the observations and the details of data reduction and photometry. Section~\ref{sec:TRGB} reports the distance to NGC 205 calculated using the tip of the red giant branch method. Section~\ref{sec:variable stars} discusses the identification and classification of variable stars and the determination of their characteristic parameters. Section~\ref{sec:simulations} describes simulations performed to assess the completeness and reliability of our classification process. Section~\ref{sec:results} presents our analysis of the catalogue, including the star formation history of NGC 205, and section~\ref{sec:conc} presents our conclusions. The catalogue itself is presented in Appendix \ref{app:catalogue}.

\section{Photometric Data} \label{sec:data}

\begin{figure}
    \centering
    \includegraphics[width=1\linewidth]{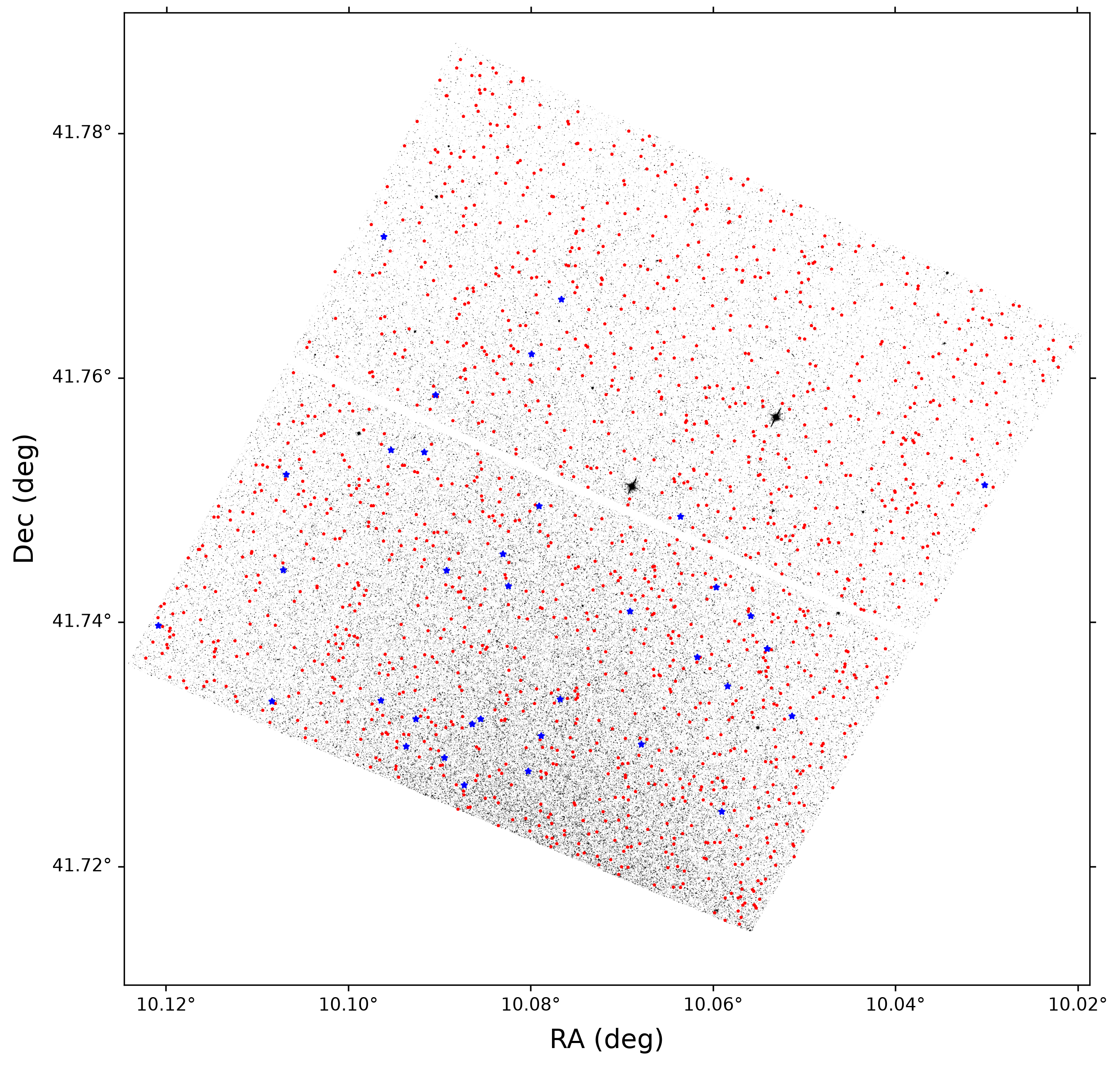}
    \caption{Spatial distribution of the variables identified in NGC 205, overplotted on an F606W image of the ACS/WFC field. RRLs are shown in red and ACs in blue.}
    \label{fig:ACS}
\end{figure}

The data used in this work were obtained as part of the NASA \emph{HST} programme GO-15902 (PI Weisz) and consist of imaging data taken in October 2020 with the Advanced Camera for Surveys Wide Field Channel (ACS/WFC) in the F606W and F814W filters. The images were downloaded from the Barbara A. Mikulski Archive for Space Telescopes (MAST). The field lies ${\sim}$4$'$ from the centre of NGC 205, and is shown in Fig.~\ref{fig:ACS} with the identified RRLs (red) and ACs (blue) overplotted, illustrating the spatial distribution of the identified variable stars.

There are 11 exposures in each band, collected in three groups over the span of 3.1 days. The first two groups consist of four images per band collected over ${\sim}$5 hr each, separated by 1.8 days. The third consists of three images per band collected over ${\sim}$2 hr, 0.8 days after the second. Within each group the two filters were interleaved. Ten of the images in each band have exposure times ranging from 868 to 1260 s; the eleventh is a 60 s frame. These short frames were included in the photometric reduction, but their depth is considerably shallower than that of the long exposures. For the variable stars in our sample, the mean photometric uncertainty in these frames is roughly an order of magnitude larger than in the long exposures (a factor of 7.3 in F606W and 9.6 in F814W), which left us with at most ten usable epochs per band. 

The data reduction and calibration follow the same procedure explained in \citet{Monelli-2010}. Accordingly, we adopted the original \textsc{flc} images provided by the \emph{HST} calibration on-the-fly reprocessing pipeline rather than the processed drizzled products. The pixel-area mask was therefore applied to each image to compensate for the different area of sky projected onto individual pixels. Photometric reduction was performed using the \textsc{daophot}\,\textsc{iv}/\textsc{allframe} package \citep{Stetson-1994}, also following the prescriptions of \citet{Monelli-2010}. Photometric calibration was performed using the standard recipe provided by the instrument documentation and updated zero points. 

\section{Distance Determination} \label{sec:TRGB}

In order to  plot the period-luminosity relation used for identification and classification of variable stars, as well as to calculate the metallicity of RR Lyrae stars using the PLZ, we must obtain an accurate distance to NGC 205. The method used in this paper, the tip of the red giant branch (TRGB) method, has been demonstrated over several decades to be a robust distance indicator (see \citealt{lee-1993,jacobs-2009,jang-2017}). \citet{lee-1993} in particular found that the distance to 12 Local Group galaxies obtained using the TRGB method showed good agreement with distances found using Cepheid and RRL variables. In order to obtain the distance moduli in this project, we use the QT calibration for the F606W and F814W filters provided by \citet{jang-2017} which is of the form 
\begin{equation}
    QT = F814W_{0} - \alpha(Color - \gamma)^{2} - \beta(Color - \gamma) \ ,
\end{equation}
where $\alpha = 0.159 \pm 0.010$, $\beta = -0.047 \pm 0.020 $, and $\gamma = 1.1$ are coefficients determined by the authors, and $ F814W_0 = -4.015 \pm 0.056 $ is the zero point in the F814W band obtained using NGC 4258 and the LMC as distance anchors. QT is a quadratic form of the TRGB magnitude corrected for the colour dependence of the F814W (or I) magnitude TRGB. The calculated QT value is then added to the TRGB magnitude in the QT system in order to find the distance modulus:
\begin{equation}
    (m-M)_{0} = QT_{\mathrm{RGB}} + M_{\mathrm{QT,TRGB}} \ .
\end{equation}

\begin{figure}
     \centering
     \includegraphics[width=1\linewidth]{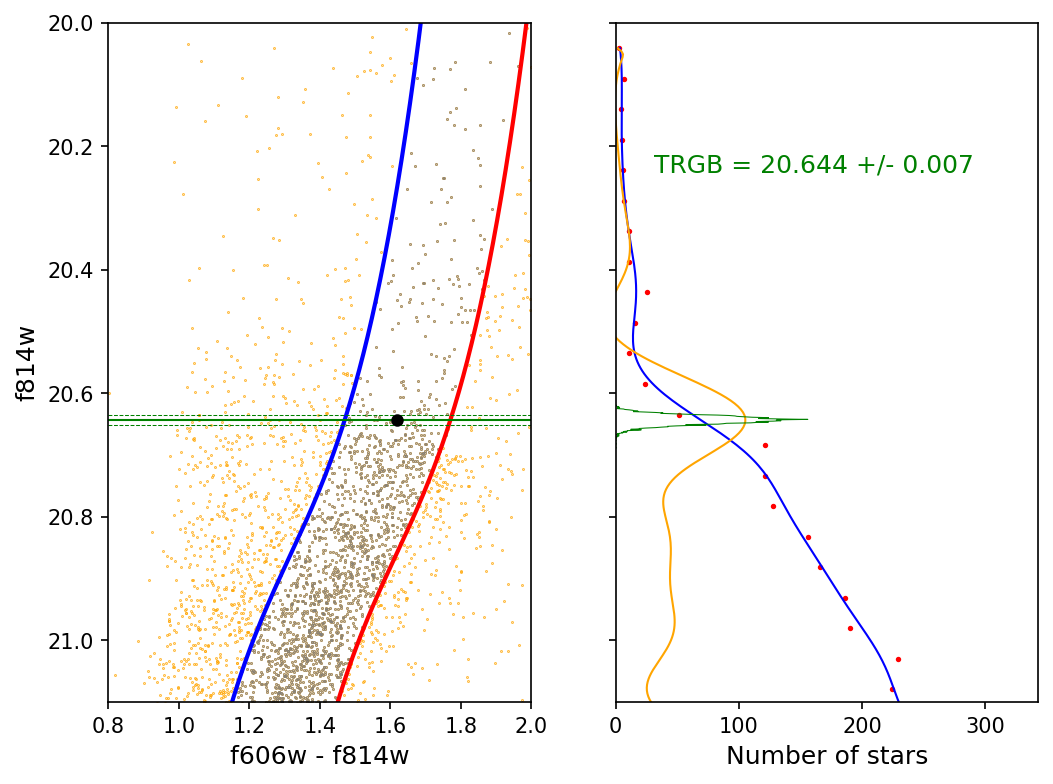}
     \caption{Determination of the TRGB in NGC 205. Left: colour-magnitude diagram in the region of the tip, with the colour range used for the measurement bounded by the blue and red lines. Right: the F814W luminosity function (red points) with its GLOESS-smoothed form (blue) and derivative (orange); the Gaussian fit to the peak of the derivative (green) gives $F814W({\rm TRGB}) = 20.64 \pm 0.01$ mag, marked by the green horizontal line in the left panel.}
     \label{fig:TRGB}
 \end{figure}

 To find the tip of the RGB, a \textsc{python} implementation of the Sobel filter method is used, which involves fitting a luminosity function (LF) to the stars near the top of the RGB and smoothing it with the GLOESS (Gaussian-windowed, Locally Weighted Scatterplot Smoothing) algorithm. A Gaussian function fit is used to find the peak of the LF derivative, which is taken to be the TRGB. The magnitudes used here have been dereddened using the Galactic extinction values for the ACS/WFC F606W and F814W bands taken from the NASA/IPAC Extragalactic Database (NED), and are 0.153 mag in F606W and 0.095 mag in F814W \citep{schlafly-2011}. This process is demonstrated in Fig.~\ref{fig:TRGB}. We found \(F814W(TRGB) = 20.64 \pm 0.01\) mag, corresponding to a distance modulus of 
\(\mu = 24.68 \pm 0.06\) mag, in agreement with other published values \citep{Butler-2005, savino-2022,sarajedini-2025}.
 
\section{Variable Stars} \label{sec:variable stars}

\subsection{Identification and Classification}
In order to identify candidate variables, we used a variable identification program (\textsc{Variable Identification}; \citealt{neeley-2021}) that calculates the Stetson variability index \citep{stetson-1996} for each star. A cut-off index was manually selected such that the vast majority of variables are recovered while keeping the number of stars to be manually classified manageable. This was done by creating a histogram of the log of the number of identified variables per variability index and choosing the index at which the slope of the histogram changes. We found that this occurred at a variability index of approximately 6. Applying this cut-off, along with a magnitude cut-off of approximately one magnitude below the HB to remove stars too faint to be those of interest, narrowed our list of potential variables significantly from more than 750,000 stars to 7682 candidates. 

Next, the periods of all candidate variables were found using Robust RR Lyrae light curve FITting (\textsc{rrfit}; \citealt{yang-2011}), a method written in \textsc{fortran} that implements a robust genetic algorithm known as \textsc{pikaia} to find the best fitting model parameters via an artificial intelligence random mesh searching technique. \textsc{rrfit} simultaneously fits two light curves obtained in different bands using 25 RRL templates from \citet{layden-1998} and \citet{kovacs-2006}. As such, it provides a consistent period in the two bands, as well as estimates of intensity-mean magnitude and amplitude in each band. The program accepts a range of periods and amplitudes over which to search. For this work, we used a period range of 0.2 to 1.2 d, and an amplitude of 0.2 to 1.5 mag, which encompasses the typical values for RRL stars.  

Finally, each candidate was examined manually to determine: (a) whether it is a true variable rather than photometric noise imitating variability, (b) its variable class, and (c) its pulsation mode (fundamental versus first-overtone). This process uses software written in \textsc{python} (\textsc{Variable Classification}; \citealt{neeley-2021}) that displays plots and images useful for classifying variable stars. Of particular note, the program displays the colour-magnitude diagram (CMD) of the galaxy along with the position of the candidate, an image of the candidate from the \emph{HST} photometry, its template-fit light curve from \textsc{rrfit}, its position on a period-luminosity (PL) plot, and its position on a Bailey diagram (a period-amplitude plot). The candidate's position on the CMD provides a quick initial determination of whether the candidate is a variable of interest, as any variable stars relevant to this project are located along the instability strip. Of course, a star can lie within the instability strip without necessarily being a variable, so the light curve was then inspected to determine whether the template was a good fit to the data and whether the apparent variability could simply be caused by large photometric errors. The photometric image aids in this judgement, as it is often evident when the photometry is poor (e.g. there may be a bright star nearby contaminating the candidate’s photometry, or the candidate may not be a star at all). 

Once a candidate was identified as a variable star, the PL plot was used to classify it by type, as different variable classes occupy different regions of the plot. To aid in this, the plot has the LMC relations of \citet{neeley-2021} from the OGLE-IV catalogues \citep{Soszynski-2015a,Soszynski-2015b,Soszynski-2016,Soszynski-2018} for RRLs, Classical Cepheids, Anomalous Cepheids, and Type II Cepheids. We further verified that the candidate’s position on the CMD was consistent with its classification from the PL plot (e.g. RRLs should lie along the HB). Occasionally, a variable star did not fit neatly into any of these classes and was labelled as an unclassified variable. In rare cases, the classification was ambiguous between an RRL and an AC, such that the star could be confidently identified as belonging to one of these two classes, but could not be definitively assigned to either; these were classified as RRL/AC. For stars that could be classified, the pulsation mode was determined via the PL plot, as fundamental and first-overtone pulsators occupy slightly different regions, with first-overtone pulsators falling slightly above fundamental pulsators. In the case of RRL stars, the pulsation mode was verified using the Bailey diagram, as RRab and RRc stars occupy distinct regions of this diagram as well. The Bailey diagram, along with the shape of the light curve (symmetric versus asymmetric), can also help differentiate between RRab and RRc when the PL diagram is inconclusive. This distinction is of particular importance for RRL stars, as the pulsation mode affects properties relevant to our future analysis of the galaxy’s chemical enrichment history. The results of this process, including the PL plot and Bailey diagram, are further discussed in Section \ref{subsec:variable results}.

\subsection{Alias Removal} \label{sec:alias}

Since only ten data points are available for each star, period aliases were very common (occurring in approximately one third of all variables), and care was taken to identify the correct period. Often, potential aliasing was apparent:  stars would display the characteristics of one pulsation mode, but have a pulsation period corresponding to another. Examples include an RRc template fit (sinusoidal) on a long-period RRL, an asymmetric light curve on a short-period RRL, a large-amplitude RRL paired with a short period, or a particularly red/blue star with a short/long period. In such cases, the range of pulsation periods and/or templates was restricted in \textsc{rrfit} to ensure an appropriate template shape for the corresponding period. When it was not immediately clear whether a star was an RRab or RRc (e.g. stars with small or intermediate amplitudes, where a large amplitude would otherwise strongly suggest an RRab), both fits were obtained using \textsc{rrfit} and the better fit -- determined both visually and by $\chi^2$ value (see Sections \ref{sec:MCMC} and \ref{sec: additional analysis}) -- was selected. For stars for which both fits were acceptable, both modes are included in the catalogue and identified with a flag (see Section \ref{sec:results}).  

\subsection{Final Values and Uncertainties} \label{sec:MCMC}

\begin{figure*}
    \centering
    \includegraphics[width=1\linewidth]{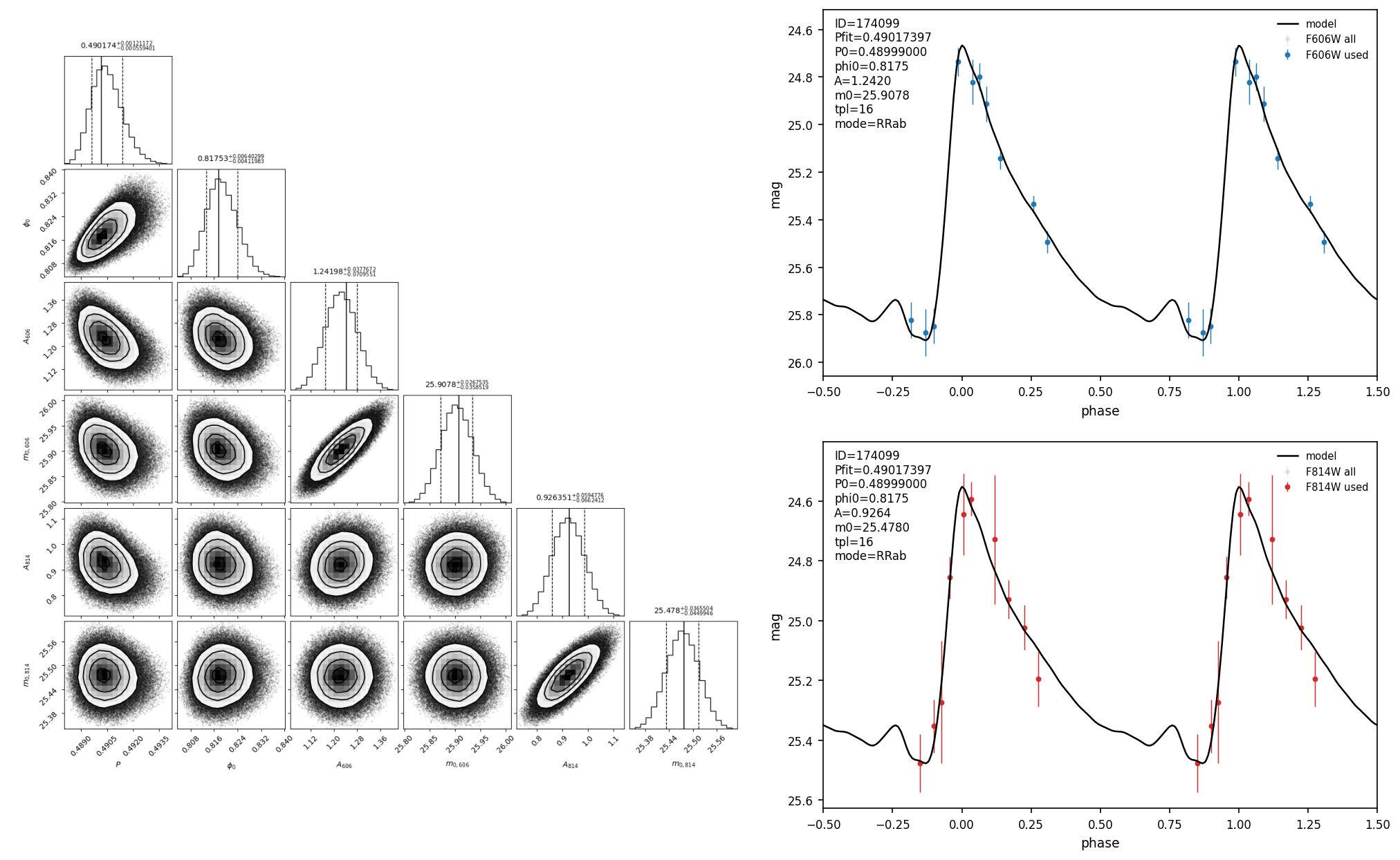}
    \caption{Posterior distributions for the MCMC fit to a sample RRab star (ID 174099), shown alongside the resulting phased light curves in F606W (top right) and F814W (bottom right). In the corner plot, the solid vertical line marks the best-fit value for each parameter and the dotted lines mark the 16th and 84th percentiles of the distribution.The solid curves on the light-curve panels are the template evaluated at the best-fit parameters, and the data are repeated over two cycles for clarity. Note that the magnitude parameters shown are peak magnitudes; the intensity-mean magnitudes reported in the catalogue are derived from these as described in Section \ref{sec:MCMC}.}
    \label{fig:MCMC}
\end{figure*}

To obtain final values and uncertainties for the periods, amplitudes, and intensity-mean magnitudes, we used a Markov Chain Monte Carlo (MCMC) approach implemented in \textsc{python} with the \texttt{emcee} package \citep{Foreman-2013}. The MCMC takes the parameters from the classification program — namely the template, period, phase, F814W mean magnitude and amplitude, and F606W mean magnitude and amplitude — and uses them to phase the data points and recreate the light curve. We adopted Gaussian priors for these parameters, using the output of the classification program as the prior mean. The standard deviation was set to 0.1 for the amplitudes and magnitudes, and to 0.05 for the period and phase; the tighter period and phase priors were chosen to reduce the risk of the MCMC converging on a period alias. Since the template is parametrised by peak magnitude rather than mean magnitude, the mean magnitude and amplitude from the classification program are first converted into a peak magnitude to serve as the initial value. The MCMC fit is then performed accordingly, and the intensity-mean magnitude reported in the catalogue is computed from the template once the best-fit parameters have been determined. 

The MCMC was run with 200 walkers for 6000 steps, with the first 1000 discarded as burn-in. The autocorrelation values ($\tau$) varied by parameter but fell between roughly 50 and 100, making the adopted burn-in at least 10$\tau$. The remaining steps were thinned by a factor of five and combined across walkers, giving 200,000 samples of the posterior. The best-fit parameters are those of the chain sample with the maximum log posterior, which combines the Gaussian priors above with a likelihood equivalent to $-\chi^2/2$. The reported uncertainties are the 99.7\% credible interval (nominally $3\sigma$) for the period and the 68\% credible interval (nominally $1\sigma$) for the amplitudes and magnitudes. The posteriors are approximately Gaussian, so the 99.7\% interval was obtained by scaling the 68\% interval by a factor of three rather than by taking percentiles of the chain directly. A corner plot of the posterior distributions for a sample RRab star is shown in Fig.~\ref{fig:MCMC}, along with the phased light curve produced using the best-fit parameters (represented by the solid line in the posterior distributions). The 16th and 84th percentiles (68\% credible interval) of the distributions are represented by the dotted lines.

\subsection{Additional Analysis} \label{sec: additional analysis}

The MCMC also served as a tiebreaker for the RRL with uncertain fits, providing a more objective measure of goodness of fit. For stars with two plausible fits, the MCMC was used to verify that the fit that looked better was also statistically preferred. We followed the prescription of \citet{Liddle-2007}, which describes the use of information criteria for model selection in astrophysics. Such criteria balance how well a model fits the data against how many free parameters it uses, so that a more complex model is only preferred if the improvement in fit outweighs the extra freedom. The Akaike information criterion (AIC) penalises each additional parameter by a fixed amount, while the Bayesian information criterion (BIC) penalises more heavily as the sample size grows. The BIC is derived under the assumption that the posteriors are near-Gaussian, while the AIC assumes the number of data points greatly exceeds the number of free parameters, which is not the case for our ten-epoch light curves. In such small-sample cases, \citet{Liddle-2007} recommends the AICc correction, which strengthens the AIC penalty and brings it closer to the BIC. 

In our case, both models have the same number of free parameters and the same number of data points, so every penalty term, including the AICc correction, is identical and cancels. The comparison is therefore the same whichever criterion is adopted, and does not depend on whether their respective assumptions are satisfied. Taking advantage of this fact, the comparison reduces to $\Delta IC = \Delta\left(-2\ln\left(\mathcal{L}_{max}\right)\right)$, where $-2\ln(\mathcal{L}_{max})$ is the $\chi^2$ of the best-fit solution. Following the Jeffreys' scale, which rates $\Delta$IC \textgreater\ 5 as strong evidence and $\Delta$IC \textgreater\ 10 as decisive evidence against the model with higher $\chi^2$, we adopted the better-fitting model only where $\Delta$IC \textgreater\ 5. In cases where  $\Delta$IC \textless\ 5, both fits are included in the catalogue and identified with a flag. 

Since we had no means of identifying aliases for ACs equivalent to that used for the RRLs, as ACs do not follow the strict relationship between mode and template shape (\citealt{Marconi-2004,Fiorentino-2012}), we used the MCMC as an independent check on the \textsc{rrfit} periods. We performed a broad MCMC fit with uniform priors spanning the same range as the \textsc{rrfit} search. Where the MCMC returned a period other than that found by \textsc{rrfit}, the fit with the lower $\chi^2$ was selected and its mode identified from its position on the PL diagram. 

\section{RRL Simulations} \label{sec:simulations}

To evaluate the completeness of our RRL search and the efficacy of our alias removal procedure, we generated a realistic sample of simulated RRL with magnitudes, photometric uncertainties, amplitudes, and periods drawn from the observed distributions in the NGC~205 data set, and analysed them with the same classification pipeline used for the observations. In addition to quantifying the success of our recovery and classification procedures, these simulations allow us to identify groups of RRL stars that cannot be reliably recovered due to limitations in our data, such as the observing cadence and number of epochs.

\subsection{Synthetic Photometry} \label{sec:synthphot}

Since the relevant parameters of an RRL star are interdependent, we used an MCMC sampler to generate a correlated set of period, F606W amplitude, and F814W magnitude for each simulated star. The likelihood function comprised two terms, one for the magnitude and one for the amplitude, each based on an empirical relation. For the magnitude, we used the F814W PL relations of \citet{neeley-2021}, derived from the OGLE-IV LMC catalogues \citep{Soszynski-2016} and shifted to the distance of NGC 205 using the distance modulus from Section \ref{sec:TRGB}. For the amplitude, RRab stars were constrained by the Oosterhoff I period-amplitude relation of \citet{Fabrizio-2019}. Since no equivalent Oosterhoff relation exists for RRc stars, we assumed a Gaussian amplitude distribution with the mean and standard deviation derived from the observed RRc sample.

The priors were chosen to reproduce the observed density structure of the Bailey diagram. For RRab stars, we constructed a two-dimensional kernel density estimate (KDE) from the observed periods and F606W amplitudes. This preserves the strong correlation between period and amplitude, with longer-period RRab stars typically having lower amplitudes, which would not be captured by independent one-dimensional priors. For RRc stars, a one-dimensional KDE on $\log P$ was sufficient, as they do not exhibit a significant period-amplitude correlation. Instead, RRc occupy a compact region of the Bailey diagram with a nearly constant amplitude distribution, which is adequately captured by the Gaussian amplitude term in the likelihood function.

The MCMC was run with 50 walkers for 3000 steps, with the first 1000 discarded as burn-in and the remainder thinned by a factor of ten; the first 1000 samples of the resulting chain were adopted as the simulated population. F606W intensity-mean magnitudes and F814W amplitudes were not sampled directly by the MCMC. Instead, each was derived from its counterpart in the other band: the F606W magnitude from the F814W magnitude using the ratio $\langle m_{606} \rangle/\langle m_{814} \rangle$, and the F814W amplitude from the F606W amplitude using the ratio $A_{814}/A_{606}$. Both ratios were obtained by averaging the per-star values measured from the observed NGC 205 RRL sample, and were computed separately for RRab and RRc.

Once the stellar parameters were generated, each simulated star was assigned a light curve template based on its period. RRab stars were assigned one of ten templates according to their period, each template selected as the most common among observed RRab stars in the corresponding period bin, while all RRc stars were assigned a sinusoidal template. Synthetic photometry was generated by sampling each star's template light curve at the actual \emph{HST} observation epochs (10 in each band), after assigning each simulated star a random epoch of maximum light. The template magnitudes at each epoch were obtained by interpolating the template light curves to the corresponding pulsation phases. Gaussian photometric noise was added to each measurement, with standard deviations of 0.070 and 0.083 mag for RRab stars in F606W and F814W respectively, and 0.064 and 0.085 mag for RRc stars, based on the median photometric errors for the observed RRL stars. The resulting light curves were written in the same format as the \emph{HST} photometry and processed through the same period-finding and classification pipeline used for the real data. In total, synthetic photometric data were generated for 1000 RRab and 1000 RRc stars.

\subsection{Classification} \label{sec:class}

Once the synthetic RRL stars were generated, they were processed through the same classification pipeline described in section \ref{sec:variable stars}. We first calculated their variability index and removed all stars with an index below 6, then used \textsc{rrfit} to determine an initial period and our classification software to identify any abnormal fits. Next, we applied our alias-removal procedure to correct these abnormal stars, which were identified by suspicious combinations of parameters (e.g. unusually large amplitudes for c-type periods, or a mismatch between the period and the assigned template type), before using our MCMC software to obtain final periods and uncertainties.

\begin{figure*}
    \centering
    \includegraphics[width=1\linewidth]{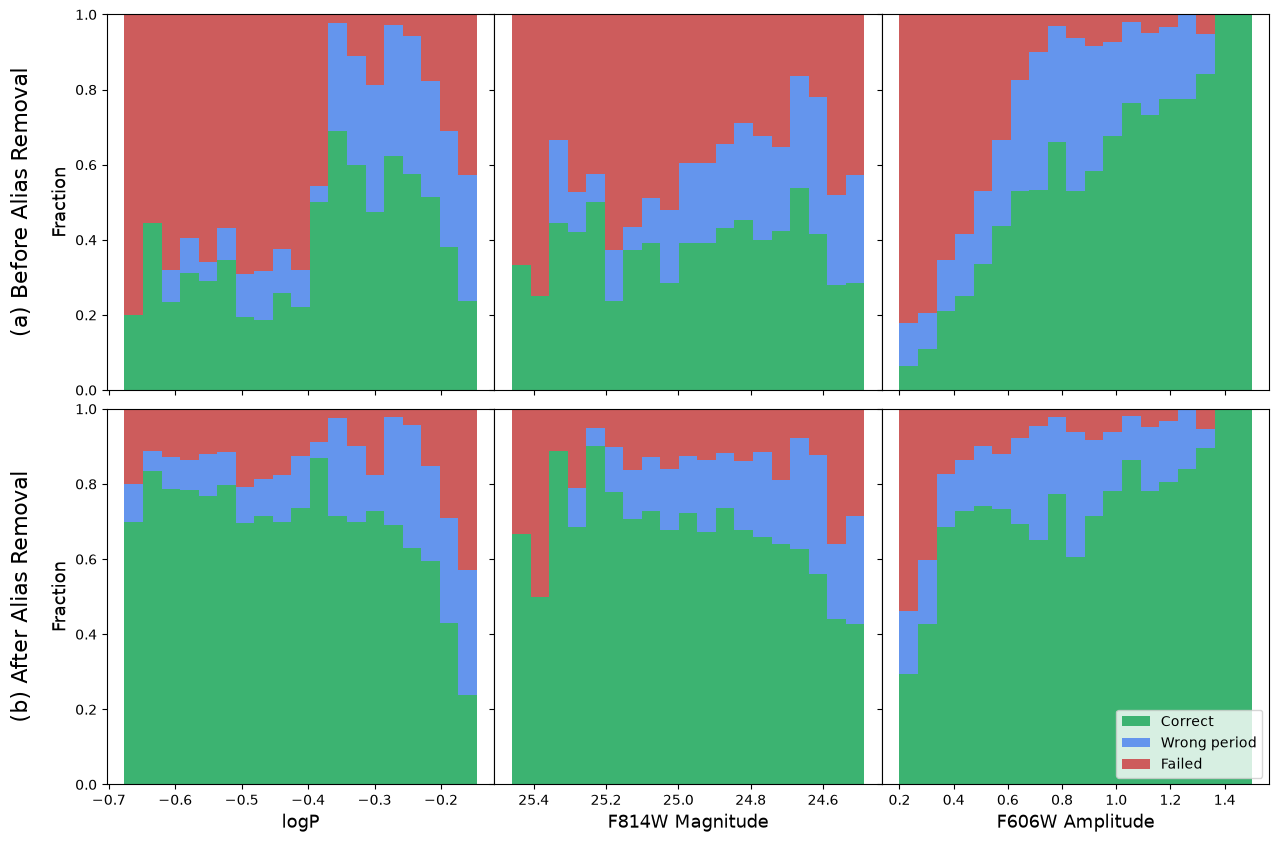}
    \caption{Recovery of the simulated RRLs as a function of input period, F814W magnitude, and F606W amplitude, before (a) and after (b) alias removal. In each bin the stars are divided into those correctly recovered in both period and mode (green), those with the correct mode but a period more than 1 per cent from the input value (blue), and those counted as failed (red), comprising stars that did not pass the variability index cut and stars assigned the wrong mode. The alias-removal procedure improves recovery across the full range of all three parameters, with the largest improvements at short periods and low amplitudes.}
    \label{fig:complete01}
\end{figure*}

Of the 2000 simulated stars, 241 (12 per cent) failed the variability index cut. Amplitude was the strongest predictor of failure, with lower-amplitude stars more likely to fail, while failures occurred across the full period and magnitude range, as shown in Fig.~\ref{fig:complete01}. Inspection of the light curves showed that even stars with intrinsically large amplitudes could fail when the observations sampled only a small portion of the pulsation cycle, most commonly near minimum light. We found no significant difference between the failure rates of RRab and RRc, at approximately 12 per cent each.  

At the initial period-finding stage the pipeline has not yet assigned a pulsation mode, so we take the fitted template shape as a proxy: a star fitted with a c-type template is treated as RRc, and one fitted with an ab-type template as RRab. A star can therefore be counted as having the wrong mode even when its period is correct. We adopt this definition because it reflects what is available in the real data, where the input period is unknown and a mismatch between template and period is itself the signal that the period may be an alias. Such a star is flagged for alias removal regardless of which of the two is at fault, and it is only during that process that it is established which is wrong. After alias removal the mode is the one assigned during classification, and this ambiguity no longer applies. We further count these stars as failed in the completeness analysis, rather than merely misclassified, since a star assigned the wrong mode is missing from the sample it belongs to and actively contaminates the other. This matters most for the mode-based quantities, such as the relative numbers of RRab and RRc, and for our upcoming PLZ analysis, which treats the two modes separately and would be biased by an RRc entering the RRab sample. 

During the initial period-finding stage, 1044 of the remaining 1759 stars (59.3 per cent) recovered the input period to within 1 per cent, with 799 (45.4 per cent) of these also fitted with a template appropriate to the assigned mode. Allowing agreement to within 5 per cent of the input period, these values increase to 1340 (76.2 per cent) and 1017 (57.8 per cent) respectively. As with the variability index, amplitude affected the success of this stage, with \textsc{rrfit} more likely to recover the correct period for higher-amplitude stars. In contrast to the variability index stage, however, the pulsation period also influenced the success of the initial period determination, with longer-period stars more likely to be recovered correctly (Fig.~\ref{fig:complete01}a). These trends are reflected in the by-mode recovery statistics. While RRab and RRc stars failed the variability index cut at nearly identical rates, \textsc{rrfit} performed substantially better for RRab stars in the initial period search: within 5 per cent of the input period, 74.0 per cent of RRab were recovered correctly, compared with 27.7 per cent of RRc. Furthermore, RRab were almost always correctly fitted with an RRab template (98.6 per cent of the time), whereas more than half (52.2 per cent) of the RRc stars were initially assigned an RRab template.

The alias-removal procedure produced a substantial improvement in the recovery rate across periods, amplitudes, and magnitudes (Fig.~\ref{fig:complete01}b). After this step, 1378 of the remaining 1759 stars (78.3 per cent) recovered the input period to within 1 per cent, and 1688 (96 per cent) were recovered within 5 per cent. This procedure also removed some, but not all, of the prior discrepancy between the two modes: 98.5 per cent of the RRab had correctly recovered periods within 5 per cent of the input period compared with 93.4 per cent of the RRc. The complete classification outcomes by mode, tolerance, and stage are given in Table \ref{tab:classification_outcomes}. Listed are the percentage of RRLs correctly classified (correct mode and period), recovered with the wrong period, or failed -- either through an incorrect mode or the variability index cut. Note that the percentages in the table are relative to the full simulated sample (1000 of each type), whereas those quoted in this paragraph exclude the stars that failed the variability index cut.

The simulations also allow us to assess how well the photometric parameters were recovered. The F606W amplitude, which sets a star's position on the Bailey diagram, was recovered less reliably, with 48.4 per cent of amplitudes within 0.05 mag of the input value and 77.5 per cent within 0.1 mag. However, recovery of the F814W magnitude, which is used in the CMD and PL relation and will be used in the PLZ analysis in the next paper in this series, was considerably better. For this parameter, 86.2 per cent were recovered within 0.05 mag and 98.8 per cent within 0.1 mag, the latter comparable to the intrinsic scatter of the PL relation.

Table \ref{tab:recovery} provides a more detailed breakdown of the alias-removal process, including the aforementioned case prior to alias removal, where the mode is inferred from the fitted template and a star can have the correct period but the wrong mode. After alias removal this category disappears entirely, since we only fit templates appropriate to the recovered period. The statistics shown in this table are for all RRL, including those that failed the variability index cut, since we were also interested in how the procedure performs on low-amplitude stars (depending on factors such as the size of the error bars, some of these could pass the cut in real data). Eleven RRab were excluded entirely: their measurements clustered near minimum light and did not resemble variables, and they would not have passed either the variability index cut or manual inspection of the template fit. Percentages are therefore relative to 989 RRab, 1000 RRc, and 1989 combined. We find that we are more likely to correctly determine the mode for RRab stars than for RRc stars, with the latter often misclassified as RRab, though the alias-removal procedure corrects most of these. Before the procedure is applied, 62.1 per cent of RRc are assigned the wrong mode, falling to 3.6 per cent afterwards, compared to 2.2 per cent and none for RRab stars, respectively. The RRc stars are also more likely to have the wrong period even when classified correctly, though not by much -- after alias removal, at the 5 per cent tolerance, 3.4 per cent of RRc stars have the wrong period compared with 1.9 per cent for RRab. Even including the low-amplitude stars, the alias-removal procedure performs well, with 95.5 per cent of all RRL correctly classified within 5 per cent of the input period.  

\begin{table}
\centering
\caption{Classification outcomes for the simulated RRL sample at two stages of the pipeline: before alias removal refers to the initial period-finding stage (\textsc{rrfit}), and after alias removal to the final classification. Results are given at two period tolerances. A star's classification is correct if both its mode and its period (within the given tolerance of the input period) are recovered. A star is classified as having the wrong period if its mode is recovered but not its period. The two failed columns give stars recovered under the wrong pulsation mode (mode) and stars not recovered as variables due to the variability index cut-off (VI). Percentages are relative to the number of stars of each type (1000).}
\label{tab:classification_outcomes}
\setlength{\tabcolsep}{6pt}
\begin{tabular}{llcccc}
\hline
\hline
\multirow{2}{*}{Stage} & \multirow{2}{*}{Type} & \multirow{2}{*}{Correct} & \multirow{2}{*}{Wrong Period} & \multicolumn{2}{c}{Failed} \\
 & & & & Mode & VI \\
\hline
\multicolumn{6}{l}{\textit{1\% Period Tolerance:} } \\
\hline
Before Alias & All  & 40.0\% & 21.2\% & 26.8\% & 12.0\% \\
Removal & RRab & 54.1\% & 32.7\% & 1.4\% & 11.8\% \\
       & RRc  & 25.8\% & 9.7\% & 52.2\% & 12.3\% \\
\hline
After Alias & All  & 68.9\% & 17.4\% & 1.7\% & 12.0\% \\
Removal & RRab & 63.4\% & 24.8\% & 0.0\% & 11.8\% \\
       & RRc  & 74.4\% & 10.0\% & 3.3\% & 12.3\% \\
\hline
\multicolumn{6}{l}{\textit{5\% Period Tolerance:}} \\
\hline
Before Alias & All  & 50.8\% & 10.3\% & 26.8\% & 12.0\% \\
Removal & RRab & 74.0\% & 12.8\% & 1.4\% & 11.8\% \\
       & RRc  & 27.7\% & 7.8\% & 52.2\% & 12.3\% \\
\hline
After Alias & All  & 84.4\% & 1.9\% & 1.7\% & 12.0\% \\
Removal & RRab & 86.9\% & 1.3\% & 0.0\% & 11.8\% \\
       & RRc  & 81.9\% & 2.5\% & 3.3\% & 12.3\% \\
\hline
\hline
\end{tabular}
\end{table}

\begin{table*}
\centering
\caption{Period and mode recovery rates for simulated RRL stars, before and after alias removal. A period is considered recovered if it falls within the given percentage of the input period. Percentages are relative to 989 RRab, 1000 RRc, and 1989 combined.}
\label{tab:recovery}
\begin{tabular}{llcccc}
\hline
\hline
 & & \multicolumn{2}{c}{Correct mode} & \multicolumn{2}{c}{Incorrect mode} \\
Stage & Type & Correct $P$ & Incorrect $P$ & Correct $P$ & Incorrect $P$ \\
\hline
\multicolumn{6}{l}{\textit{1\% Period Tolerance:}} \\
\hline
Before Alias Removal & RRab     & 56.2\% & 41.6\% & 0.4\% & 1.8\% \\
            & RRc      & 27.3\% & 10.6\% & 27.4\% & 34.7\% \\
            & Combined & 41.7\% & 26.1\% & 13.9\% & 18.2\% \\
\hline
After Alias Removal & RRab     & 67.7\% & 32.3\% & 0.0\% & 0.0\% \\
                     & RRc      & 83.2\% & 13.2\% & 0.0\% & 3.6\% \\
                     & Combined & 75.5\% & 22.7\% & 0.0\% & 1.8\% \\
\hline
\multicolumn{6}{l}{\textit{2\% Period Tolerance:}} \\
\hline
Before Alias Removal & RRab     & 70.6\% & 27.2\% & 0.7\% & 1.5\% \\
            & RRc      & 29.4\% & 8.5\% & 35.1\% & 27.0\% \\
            & Combined & 50.0\% & 17.8\% & 17.9\% & 14.2\% \\
\hline
After Alias Removal & RRab     & 86.8\% & 13.2\% & 0.0\% & 0.0\% \\
                     & RRc      & 92.6\% & 3.8\% & 0.0\% & 3.6\% \\
                     & Combined & 89.7\% & 8.5\% & 0.0\% & 1.8\% \\
\hline
\multicolumn{6}{l}{\textit{5\% Period Tolerance:}} \\
\hline
Before Alias Removal & RRab     & 80.4\% & 17.4\% & 1.0\% & 1.2\% \\
            & RRc      & 29.4\% & 8.5\% & 36.5\% & 25.6\% \\
            & Combined & 54.9\% & 13.0\% & 18.8\% & 13.4\% \\
\hline
After Alias Removal & RRab     & 98.1\% & 1.9\% & 0.0\% & 0.0\% \\
                     & RRc      & 93.0\% & 3.4\% & 0.0\% & 3.6\% \\
                     & Combined & 95.5\% & 2.7\% & 0.0\% & 1.8\% \\
\hline
\hline
\end{tabular}
\end{table*}

\section{Results and Discussion} \label{sec:results}

\subsection{Variable Star Population of NGC 205}\label{subsec:variable results}

\begin{figure}
    \centering
    \includegraphics[width=1\linewidth]{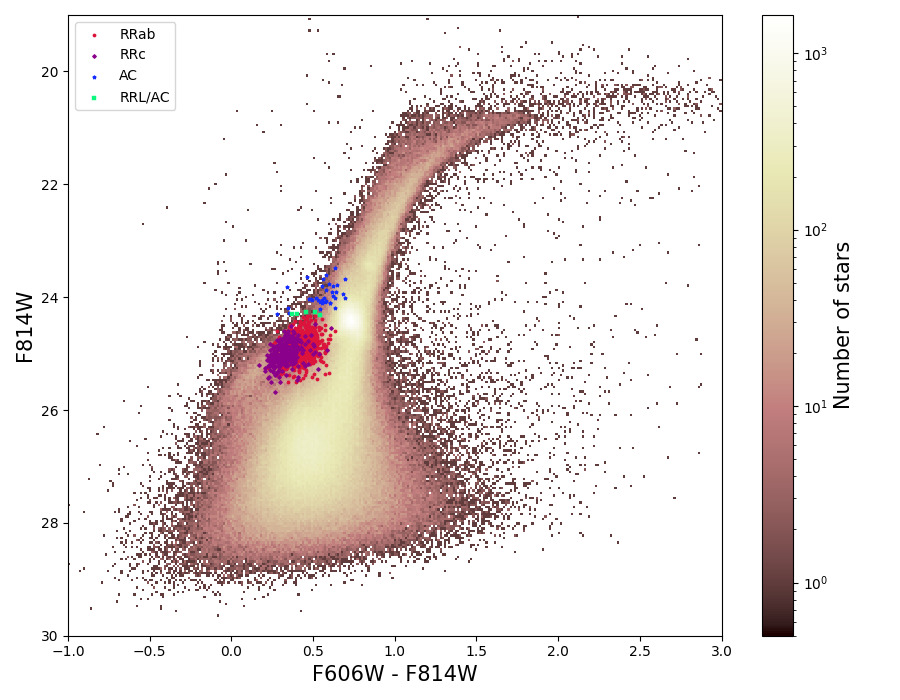}
    \caption{Colour-magnitude diagram of NGC 205, with the density of stars shown on a logarithmic scale. The identified variables are overplotted: RRab (red), RRc (purple), AC (blue), and the five stars that could not be distinguished between RRL and AC (green).}
    \label{fig:CMD}
\end{figure}

\begin{figure}
    \centering
    \includegraphics[width=1\linewidth]{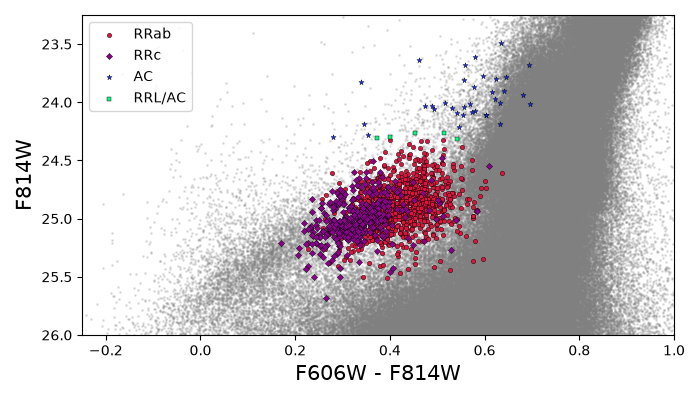}
    \caption{Colour-magnitude diagram of NGC 205 zoomed in on the horizontal branch, with non-variable stars shown in grey. The RRab and RRc separate in colour along the instability strip, while the AC lie above the horizontal branch as expected for their higher luminosities. }
    \label{fig:HB_CMD}
\end{figure}

In total, our catalogue (Table \ref{tab:rrl_full}) contains 1441 RRL, 35 ACs, and five stars that could be identified as an RRL or AC but not distinguished between the two (RRL/AC). These variables are shown on the full CMD in Fig.~\ref{fig:CMD}, and on a zoomed-in CMD focused on the HB in Fig.~\ref{fig:HB_CMD}. Stars with poor template fits or poor photometry are flagged PF in the catalogue. 

Of the 1441 RRL, 1064 are fundamental mode pulsators (RRab) and 363 are first-overtone pulsators (RRc). The remaining 14 RRL either lack an obvious mode (2; flagged UM), or have acceptable fits at two periods corresponding to both an RRab and an RRc classification (12; flagged 2F). For the latter, both solutions are included in the catalogue. The RRL pulsation periods range from 0.200 to 0.836~d, with mean periods of 0.561 and 0.317 d for the RRab and RRc stars, respectively. 

Of the 35 ACs, nine are fundamental pulsators, 25 are first-overtone pulsators, and one is of uncertain mode. Since our period search did not extend past 1.2 d, the maximum period of the detected ACs is likely determined by this imposed search limit rather than any intrinsic property of the AC population. However, the minimum period of our detected ACs was 0.326 d, near the lower limit reported for the class. For these shortest-period ACs we found no other acceptable fit within our search range, though we do not rule out that they are aliases of periods longer than 1.2 d.  Given our focus on the short-period regime, it is also unsurprising that the majority of the ACs identified are first-overtone pulsators.

%\textbf{We did not extend the search, since the AC periods do not enter our analysis.}

\begin{figure*}
    \centering
    \includegraphics[width=1\linewidth]{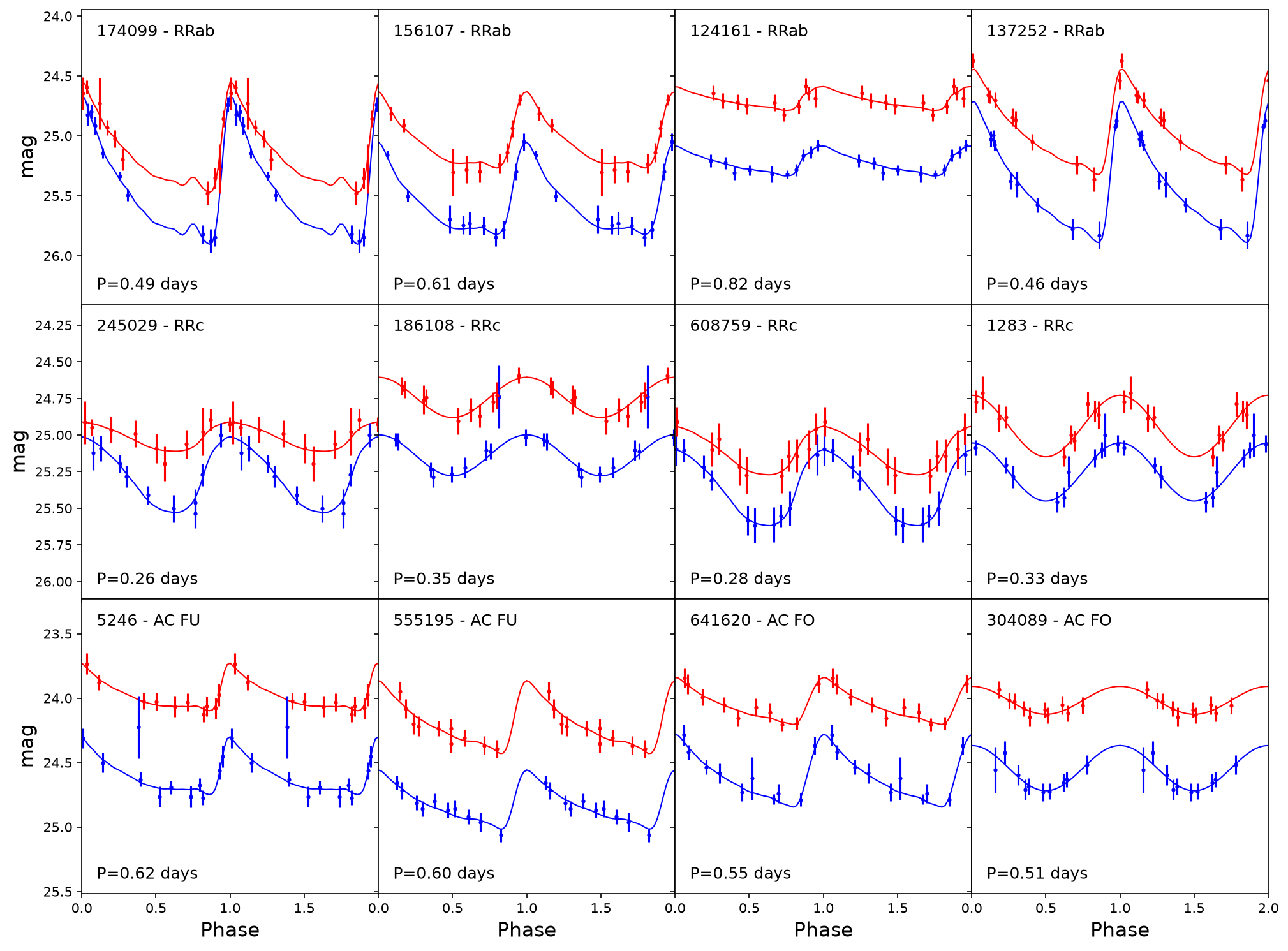}
    \caption{Phased light curves for a representative sample of variables in NGC 205: RRab (top row), RRc (middle row), and AC (bottom row), with the fundamental and first-overtone ACs labelled FU and FO. F814W and F606W data are shown in red and blue, respectively, with the corresponding template fits overplotted as solid lines in the same colours. Error bars give the photometric uncertainty on each measurement. Each panel is labelled with the star ID and fitted period, and the data are repeated over two pulsation cycles for clarity.}
    \label{fig:lc}
\end{figure*}

\begin{figure}
    \centering
    \includegraphics[width=1\linewidth]{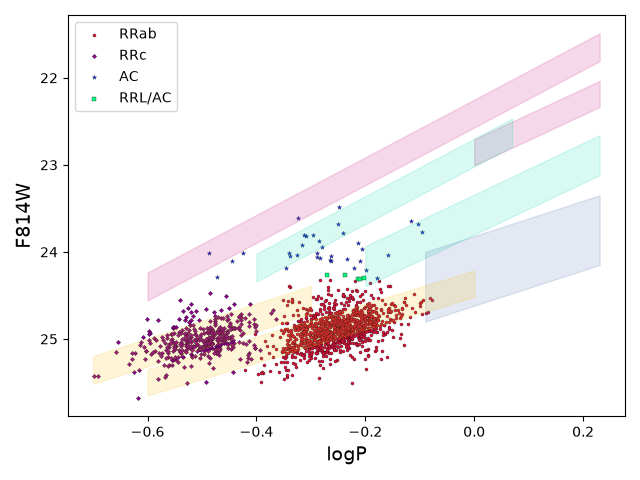}
    \caption{Period-luminosity diagram for the variables identified in NGC~205: RRab (red), RRc (purple), AC (blue), and the five stars that could not be distinguished between RRL and AC (green). The shaded regions are the LMC relations from \citet{neeley-2021}, shifted to the distance of NGC~205. These correspond to classical Cepheids (pink), anomalous Cepheids (cyan), Type II Cepheids (blue), and RR Lyrae (yellow). The relations for the classical Cepheids, anomalous Cepheids, and RR Lyrae stars are each split into fundamental and first-overtone modes, with the first-overtone relation the brighter of the two.}
    \label{fig:PL}
\end{figure}

Sample light curves for both FU and FO ACs, as well as RRab and RRc stars, can be found in Fig.~\ref{fig:lc}. The PL diagram for the RRL and AC stars is shown in Fig.~\ref{fig:PL}, which notably reveals considerable scatter in magnitude for both RRab and RRc, indicating a range of metallicities within the RRL population. A few ACs fall in the region occupied by classical Cepheids, which may reflect period aliasing or genuinely short periods; their position on the CMD, well below the classical Cepheid luminosities, confirms the AC classification.

\begin{figure}
    \centering
    \includegraphics[width=1\linewidth]{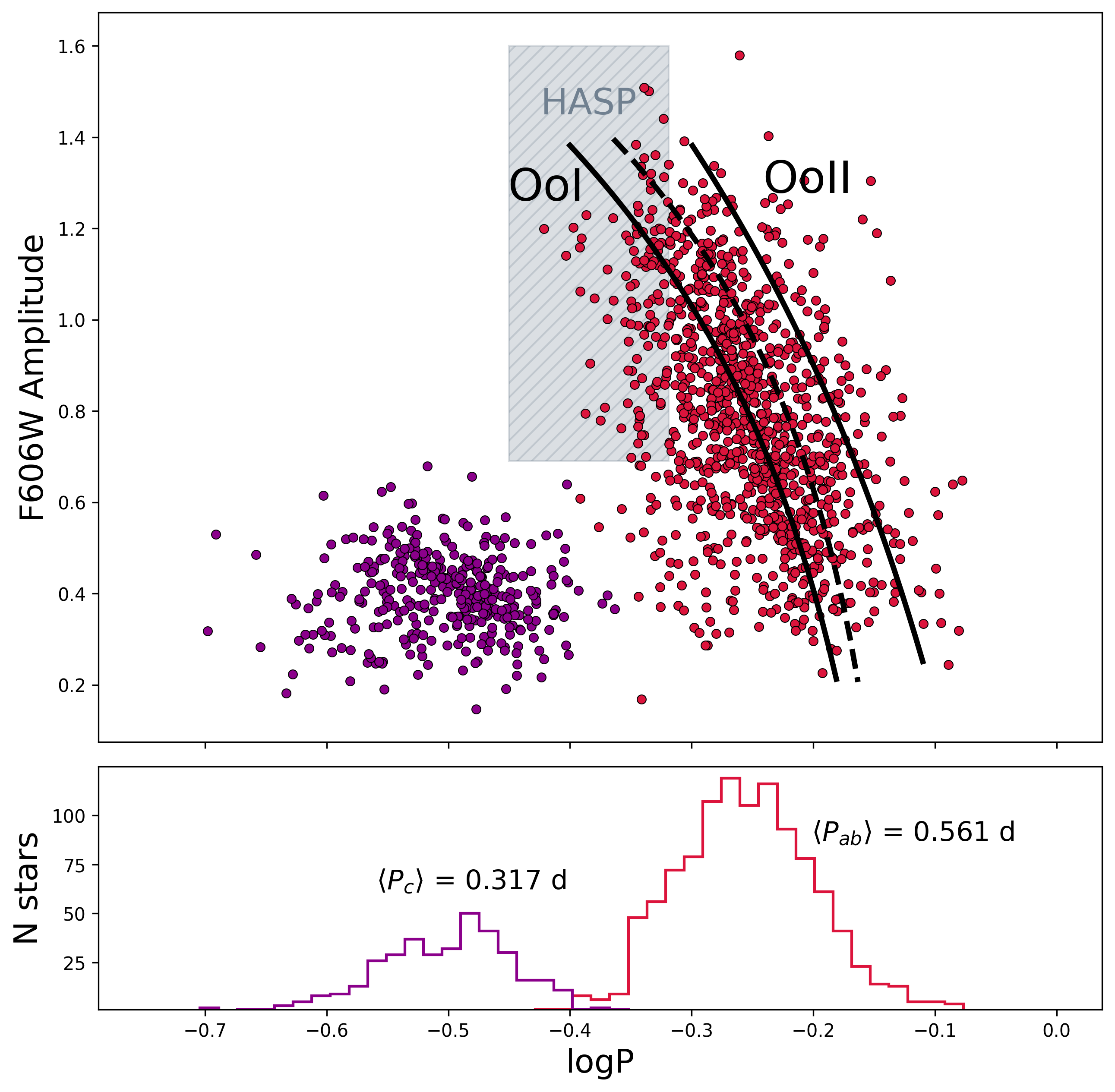}
    \caption{Top: Bailey diagram for the RR Lyrae stars in NGC 205, with RRab shown in red and RRc in purple. The solid lines mark the Oosterhoff I and Oosterhoff II relations of \citet{Fabrizio-2019}, and the dashed line the Oosterhoff intermediate relation; these were derived for the Johnson $V$ band and have been scaled to F606W assuming $A_{F606W}/A_V = 0.92$. The gray coloured region indicates the location of high-amplitude, short-period (HASP) RRab stars. Bottom: period distributions for the two modes, with mean periods as labelled.}
    \label{fig:BD}
\end{figure}

In Fig.~\ref{fig:BD}, we present the Bailey diagram as well as a histogram of the RRL period distributions. Overplotted on the RRab stars are the Oosterhoff lines as determined by \citet{Fabrizio-2019}. Since these lines were determined for the Johnson V band rather than the native \emph{HST} F606W filter, we scaled them to F606W assuming $A_{F606W}/A_V = 0.92$ (\citealt{Fiorentino-2015}, who adopted the midpoint of the 6--10 per cent range given by \citealt{Brown-2004}). As is typical of dwarf galaxies, NGC 205 appears to be an Oosterhoff intermediate system. Rather than following a single Oosterhoff locus, the RRab stars extend from short periods to the left of the OoI sequence to long periods past the OoII sequence, while also covering the intermediate region between the typical values seen in Galactic globular clusters \citep{Oosterhoff-1939}. This indicates a range of metallicities among the RRL stars, consistent with the magnitude scatter found in the PL relation. In particular, a substantial number of RRab stars lie on the short-period side of the OoI locus, which points to a significant metal-rich component within this broader distribution.

One feature of the RRab population that is clear from the Bailey diagram is the presence of high-amplitude, short-period RRab stars (HASPs), defined as fundamental-mode RRLs with $P \lesssim 0.48$ d and a V-band amplitude of at least 0.75 mag \citep{Fiorentino-2015,Fiorentino-2017}. These objects occur in stellar populations more metal-rich than $[\mathrm{Fe/H}] {\sim} -1.5$, and thus are an important signature of the fast chemical enrichment of old stars. To identify them in our sample, we applied the same amplitude scaling used for the Oosterhoff lines, resulting in a minimum amplitude of 0.69 mag in F606W. Using these criteria, we find 118 HASPs, making up 11 per cent of the RRab stars. We will examine the metallicity distribution in more detail in the next paper in this series.     

In addition to the RRL and AC populations, the catalogue contains 105 potential variable stars that could not be confidently classified (flagged as UC). These objects either occupy unusual positions on the CMD or have combinations of periods, amplitudes, and magnitudes that make their classifications uncertain. Several may be additional RRL or AC stars, but their classification is complicated by large photometric uncertainties or contamination from nearby bright or red stars, either of which can affect their measured properties and CMD positions.

For each of these variables and variable candidates, the catalogue lists the measured properties and classifications. Positions are given as right ascension and declination in degrees (J2000). Periods are in days, and the epoch of maximum light, $t_{0}$, is given as an MJD. Amplitudes and intensity-mean magnitudes are given in magnitudes for both the F606W and F814W bands. The reduced $\chi^{2}$ of the best-fit solution is also listed. Where a classification was possible, the variable type and pulsation mode are listed, with FU and FO denoting fundamental and first-overtone pulsators, respectively. Unclassified variable candidates and stars whose pulsation mode could not be determined contain a -- in the corresponding column in addition to their UC and UM flags. 

Each period, amplitude, and magnitude is accompanied by separate lower and upper uncertainties, reflecting the asymmetry of the posterior distributions described in section \ref{sec:MCMC}. As discussed in section \ref{sec:MCMC}, the amplitude and magnitude uncertainties correspond to the 68\% credible interval of the posterior, while the period uncertainties correspond to the 99.7\% credible interval; the larger interval reflects the difficulty of constraining the period with only ten epochs, and is intended to encompass the range over which an acceptable fit can be obtained.

\subsection{Star Formation History as Traced by the Variable Star Populations} \label{sec:Star Formation}

\subsubsection{RR Lyrae stars}

\begin{figure}
    \centering
    \includegraphics[width=1\linewidth]{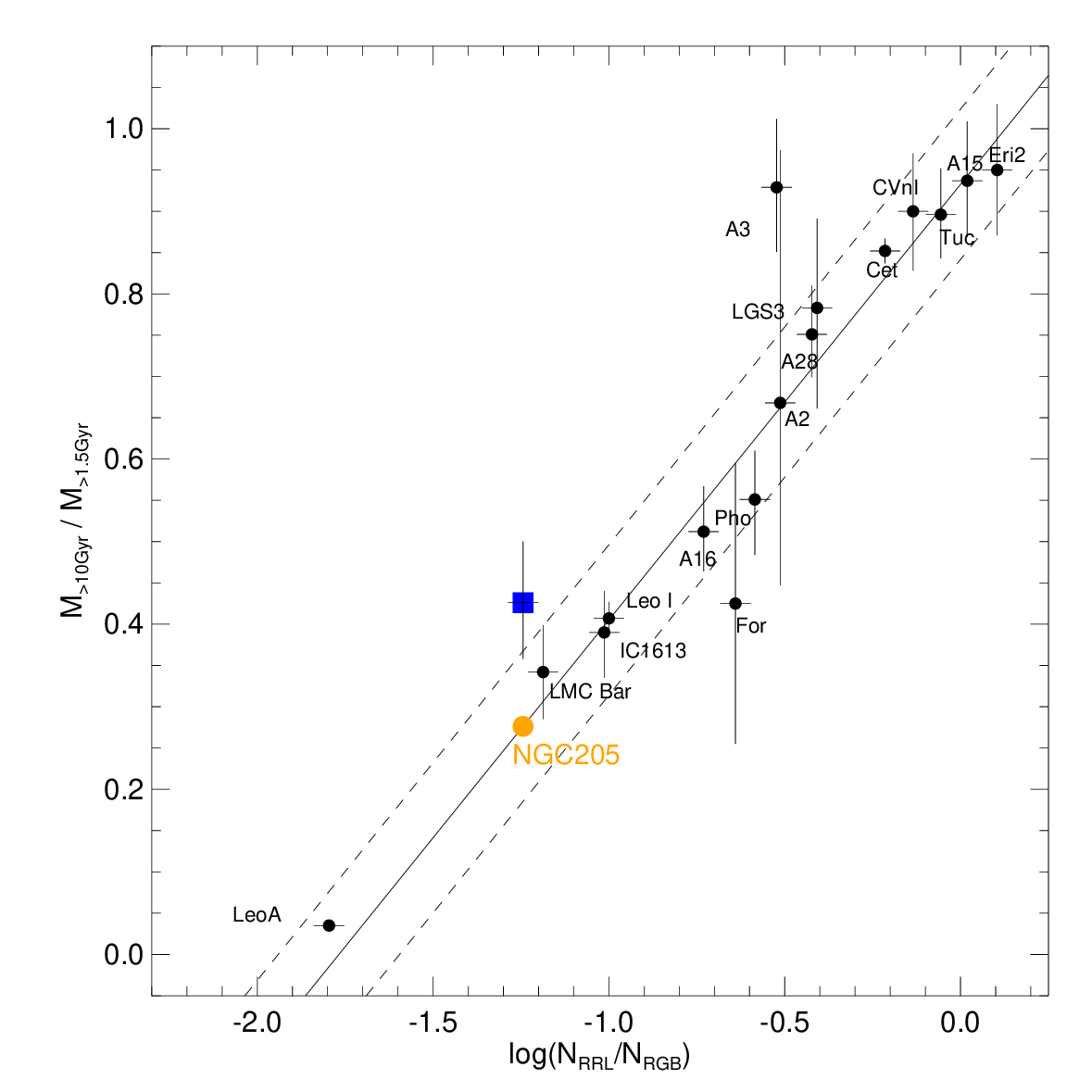}
    \caption{Correlation between the ratio of RRL to bright RGB stars and the fraction of mass formed at the earliest epochs (${>}10$ Gyr ago) to the mass formed more than 1.5 Gyr ago. Black points represent galaxies used in the calibration sample of Monelli et al. (in prep), the solid line is the best-fit relation, and the dashed lines mark the $1\sigma$ scatter. The orange circle represents the most likely value of the mass fraction given the ratio of RRL to RGB stars found in this work. The blue square shows the value derived from the \citet{savino-2025} star formation history, as described in the text.}
    \label{fig:star_formation}
\end{figure}

Fig. \ref{fig:star_formation} presents the correlation between the ratio of RRL to bright RGB stars and the fraction of mass formed at the earliest epochs (${>}10$ Gyr ago). This behaviour has been calibrated using a sample of galaxies with both variability surveys and measured star formation histories. A detailed discussion is presented in a companion paper (Monelli et al. in prep.), but we highlight the most relevant aspects here. 

The star formation histories come from deep CMDs reaching below the oldest main-sequence turn-off, allowing the mass formed at early times to be estimated directly. RRLs and stars in the top 2.5 magnitudes of the RGB are simply counted from the corresponding CMDs. They are selected because their parent stellar populations cover different age ranges: over the range of metallicity typical of dwarf galaxies, RRL stars require ages greater than ${\sim}$10 Gyr \citep{walker89, glatt08b}, whereas the upper RGB is populated by any population older than ${\sim}$1.5 Gyr \citep{Cassisi-2016}. 

Fig. \ref{fig:star_formation} clearly demonstrates that galaxies whose star formation was confined to the earliest epochs sit at the high end of the relation, while those with extended star formation sit at the low end. This is a consequence of the fact that a galaxy that formed all of its stars early on will therefore have produced both RRL and RGB stars, while one with extended star formation continues to produce RGB stars without adding to its RRL population. The ratio $\log(N_\mathrm{RRL}/N_\mathrm{RGB})$ consequently decreases as the fraction of mass formed at old epochs decreases.

We apply this method to NGC 205 as follows. From our catalogue, we find 1441 RRL stars, plus an additional 14 recovered from the Savino catalogue, giving us a total of 1455. Our completeness analysis found that 12 per cent of RRLs are missed owing to the variability index cut, which places an upper bound of 1614 on the RRL count. 

To estimate the RGB count, we defined a strip on the CMD following the ridge line of the RGB, using the same method described in Section \ref{sec:TRGB} and adopting the TRGB measured there. The colour boundaries were set by the blue and red edges, which were placed symmetrically about the ridge line and offset until the strip followed the width of the RGB while excluding likely AGB and field star contamination. The strip was then bounded in magnitude by the TRGB and TRGB+2.5. Within these limits, we obtain an RGB count of 25,572. 

To calculate the upper and lower bounds on the RGB count, we made the following assumptions:
\begin{enumerate}
    \item that the TRGB could differ from our measurement by as much as 0.1 mag in either direction; ten times its formal uncertainty.
    \item that the stars in the brightest 0.5 mag of the RGB could be either all RGB or all AGB.  
\end{enumerate}

Following these assumptions, the smallest RGB count occurs if the TRGB is 0.1 mag brighter and all stars in the brightest 0.5 mag are AGB stars, giving a lower bound of 22,848. The largest occurs if the TRGB is 0.1 mag fainter and all stars in the brightest 0.5 mag are RGB stars, giving an upper bound of 28,244.

Combining these, we find a ratio of $\log(1455/25572) = -1.24$, with upper and lower bounds of $\log(1614/22848) = -1.15$ and $\log(1455/28244) = -1.29$, adopting the largest RRL count with the smallest RGB count and vice versa.

Placing NGC 205 on this relation, represented by the orange circle, gives a mass fraction formed before 10 Gyr of approximately 0.28, with the bounds corresponding to 0.26 to 0.33.  This puts it toward the lower end of the calibrating sample, in the region occupied by the LMC fields, and well below the early-quenching dwarf spheroidals. This implies that only a modest fraction of the stellar mass in this field formed in the first few Gyr, and that star formation continued well beyond the epoch at which RRLs could still be produced.

This is broadly consistent with what is known of the galaxy from other tracers. \citet{savino-2025} derived the star formation history of NGC 205 by fitting the full CMD, and found that $\tau_{50} = 8.5$ Gyr, meaning that half the stellar mass in this field formed within the last 8.5 Gyr. The fraction formed before 10 Gyr is therefore well below half, in line with the low ratio we measure. Our RRL results support this: the position of the RRab in the Bailey diagram indicates the presence of an intermediate-to-metal-rich population, as expected if star formation continued long enough to enrich the gas.

However, while the two results agree qualitatively, the mass fraction we infer is lower than that implied by the star formation history of \citet{savino-2025}, represented in Fig. \ref{fig:star_formation} by the blue square. This difference can be reconciled in one of several ways. First, since our mass fraction is calculated from the ratio of RRL to RGB stars, a substantial incompleteness in the RRL catalogue, or an overestimate of the RGB due to contamination by AGB stars or the M31 halo, beyond what is captured by our uncertainties, would bias the inferred fraction downward. 

Another possibility is that the star formation history from \citet{savino-2025} underestimates the ancient population. Ancient star formation is least constrained at the oldest ages, since crowding and the limited signal-to-noise at the ancient main-sequence turn-off leave the oldest ages poorly sampled. If the duration of the earliest epoch is not well constrained, the division of stellar mass between the ancient- and intermediate-age populations derived from the same solution is correspondingly uncertain.

Finally, it is possible that the metallicity distribution of NGC 205 may affect the ratio itself. RRLs form only from horizontal branch stars that cross the instability strip, and the fraction that do so decreases with increasing metallicity. Consequently, if rapid early enrichment produced an ancient population with sufficiently high metallicity to populate a predominantly red horizontal branch, that population could contribute substantially to the RGB count while producing relatively few RRLs. We see some evidence for this possibility: the Bailey diagram shows RRab stars spanning the full range from metal-poor to metal-rich, the PL diagram shows substantial scatter, and the horizontal branch is notably extended in colour. Together, these point to a range of metallicities within the RRL population which we will quantify in a forthcoming paper. Such a spread indicates that star formation continued long enough for the gas to enrich appreciably, perhaps longer than the star formation history of \citet{savino-2025} suggests. This is further reinforced by the substantial HASP population, which requires a metal-rich environment to form.

\subsubsection{Anomalous Cepheids}
\begin{figure}
    \centering
    \includegraphics[trim={225pt 100pt 225pt 100pt}, clip,width=1\linewidth]{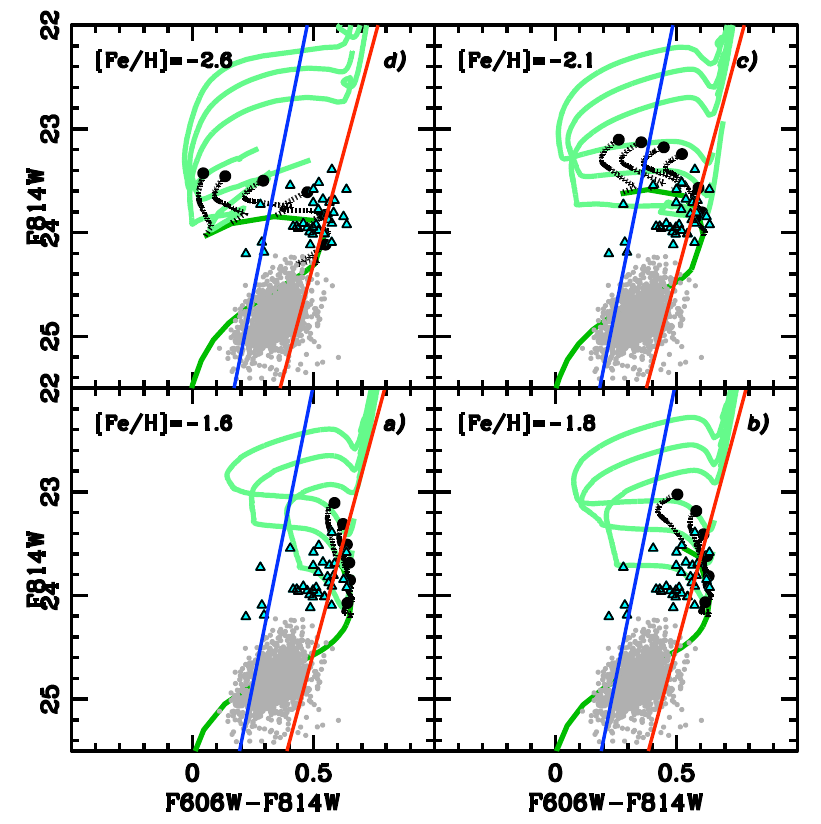}
    \caption{Zoom on the colour-magnitude plane for variable stars in NGC~205. The central helium burning loci (green lines) have been plotted for some selected metallicities (labelled in each panel). Evolutionary tracks have also been shown for stars with masses larger than 1 M$_{\odot}$ that ignite helium with a flash (dashed black lines) or quiescently (green lines). The black dots mark the end of helium burning in the centre. The instability strips have also been chosen according to theoretical models (see the text for details). RRLs are shown in grey and ACs are represented by larger triangles to highlight their agreement with theoretical models. }\label{fig:AC_CMD}
\end{figure}

Further information on the star formation history of NGC 205 can be gathered from the estimated metallicity of the AC population. Figure \ref{fig:AC_CMD} presents a comparison between the observed CMD and theoretical models from the BaSTI database \citep{hidalgo-2018}, using $\alpha$-enhanced models and fixing the distance modulus. Each panel shows a selected metallicity in the range $[\mathrm{Fe/H}] = -2.6$ to $[\mathrm{Fe/H}] = -1.6$, following the results of \citet{Bellazzini-2011}. The instability strip (IS) for RRLs is reproduced in each panel (blue and red lines), derived from the models presented in \citet{marconi-2015}. The dark green line shows the central helium burning locus for stars that experience the helium flash, from ${\sim}0.5$ to ${\sim}2.1 M_{\odot}$. For selected masses between ${\sim}1$ and ${\sim} 2 M_{\odot}$, evolutionary tracks are plotted as dotted black lines up to central helium exhaustion (black filled circles). Evolutionary tracks for more massive stars (2.2, 2.6, and 4.0 $M_{\odot}$) that ignite helium quiescently are shown as light green lines.

The agreement between the observed ACs and the models varies with metallicity. At the higher values shown in panels (a) and (b), the tracks become too red to cross the IS \citep{Fiorentino-2006}, which sets a lower limit on the metallicity of the bulk of the AC population. The models at $[\mathrm{Fe/H}] = -2.1$ and $[\mathrm{Fe/H}] = -2.6$, however, are able to reproduce the location of the faintest ACs. This points to a possible intrinsic metallicity spread within the AC population of NGC 205.

\begin{figure}
    \centering
    \includegraphics[trim={215pt 100pt 215pt 100pt}, clip, width=\linewidth]{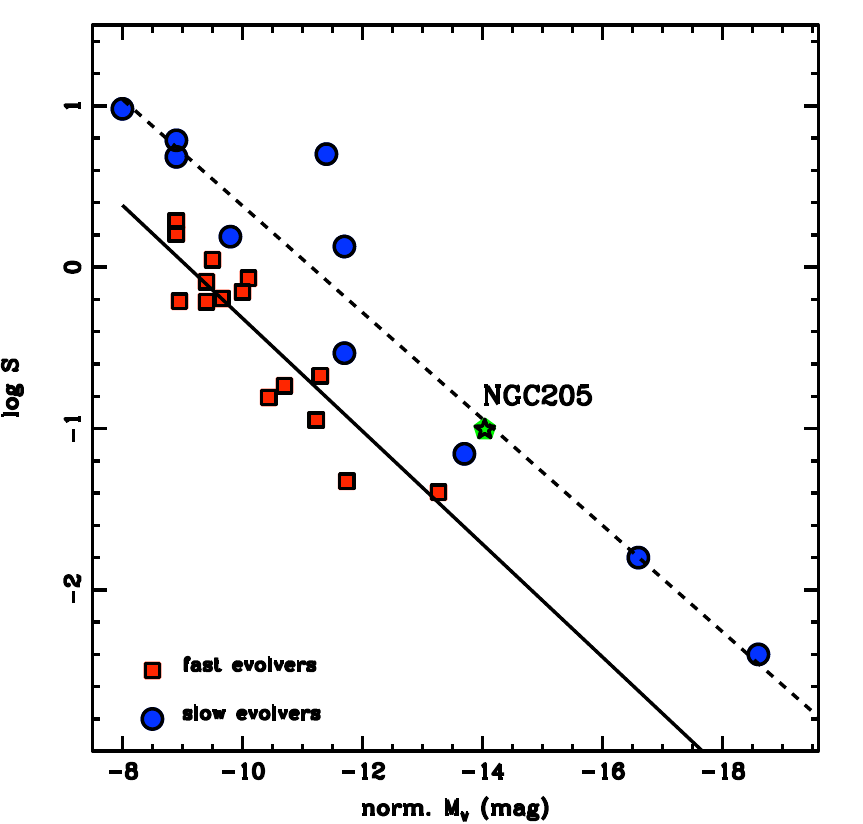}
    \caption{AC frequency as a function of the absolute visual magnitude of the host system for a sample of Local Group galaxies. The S parameter is defined as the number of AC normalised to the visual luminosity of the host galaxy, in units of $10^5 L_{\odot}$. Purely old galaxies (fast) are shown in red, while dwarfs with extended SFH (slow) are indicated by blue symbols. NGC 205 is indicated by the green star.}
    \label{fig:AC Frequency}
\end{figure}

Furthermore, the 35 identified ACs allow us to place NGC 205 on the relation between AC frequency, represented by $\log S$ (the number of ACs per $10^5~\mathrm{L}_\odot$), and host galaxy luminosity (Fig. \ref{fig:AC Frequency}). This relation was first noted by \citet{Mateo-1995} and has since been extended to a larger sample of Local Group galaxies \citep{Fiorentino-2012,Monelli-2022,Monelli-2025}. Galaxies fall along two distinct sequences: fast evolvers, which formed their stars early and then stopped \citep{Gallart-2015}, and slow evolvers, with extended star formation, which host more ACs at the same luminosity.

The reason for the two distinct relations lies in the two channels through which ACs can form. The first is the evolution of single stars of roughly 1.3--2.1 M$_\odot$, which traces a population between about 1 and 6 Gyr old \citep{Fiorentino-2006}. The second is the merger of old binary systems, linked evolutionarily to blue stragglers \citep{Renzini-1977}, which traces the old population instead. The two channels therefore sample very different ages, but they share a requirement on metallicity: in both cases the evolutionary tracks only cross the instability strip for $[\mathrm{Fe/H}] \lesssim -1.3$ \citep{Monelli-2022}, which makes ACs reliable tracers of a metal-poor component whichever channel produced them. The division into two sequences follows from which channels are available: only the binary system evolution route is viable for fast evolvers, whereas slow evolvers form ACs through both channels, and so produce more of them for a given global luminosity.

This plot places NGC 205 among the slow evolvers, indicating that both formation channels have contributed and that star formation continued well past the earliest epochs. This is consistent with the extended history found by \citet{savino-2025} from the full CMD of the same field.

Since ACs form only from metal-poor progenitors, unlike RRLs, which face no such restriction and continue to form as the galaxy enriches, metal-poor gas must have been available at intermediate ages for the single-star channel to have operated. This metal-poor gas does not appear to have been confined to one region. The spatial distribution of the ACs (Fig. \ref{fig:ACS}) is close to uniform across the field, with no concentration in any particular region, suggesting metal-poor star formation occurred throughout the galaxy rather than in a single localised episode.

\subsection{Comparison with Other Works} \label{sec:comparison}

Other papers have published similar catalogues of RRL stars in NGC~205, namely \citet[hereafter Savino]{savino-2022} and \citet[hereafter Sarajedini]{sarajedini-2025}. This catalogue is substantially more complete, with 1441 RRL found, plus an additional 35 AC and five RRL/AC, compared to 406 RRL found by Savino and 479 by Sarajedini. 

Since our analysis of the star formation history depends on the number of RRLs found, having the most complete catalogue is essential. As mentioned in Section \ref{sec:Star Formation}, we found a logarithmic ratio of $-1.24$, placing NGC 205 among galaxies with extended star formation histories such as the LMC fields. Using the RRL count from Savino or Sarajedini, the ratio becomes $-1.80$ or $-1.73$ respectively, which would place it at the extreme end of the relation alongside Leo A, a dwarf irregular that is still forming stars today. Since the ratio is interpreted as the fraction of stellar mass formed at early epochs, these ratios would imply a substantially smaller ancient population than the one we detect.

Aside from the total RRL count, we also differ on the fractions of RRc and RRab in NGC 205. While these fractions are not directly used in our analysis, the fraction of RRc relative to total RRL provides information about metallicity, with metal-rich systems containing fewer RRc than metal-poor systems. In our catalogue, we find that roughly 25 per cent of all RRL are RRc. Assuming this fraction is representative of the entire sample of RRL -- a reasonable assumption given that our completeness analysis found we were equally likely to miss RRc and RRab -- this suggests an intermediate-to-metal-rich RRL population, in agreement with our Oosterhoff population analysis discussed in Section \ref{sec:results}. It is also consistent with the extended star formation history implied by our RRL-to-RGB ratio: enrichment builds up over successive generations, so a metal-rich population is one that formed after earlier star formation had already enriched the gas. This value does not agree with either the Savino or Sarajedini catalogues, from which we calculate fractions of roughly 35 per cent and 8 per cent, respectively. If these fractions are similarly assumed to be representative, then the Savino result would suggest a more metal-poor system, while the Sarajedini result would suggest a metal-rich one. 

To ensure we have the most complete sample for the RRL-to-RGB ratio, the RRc fraction, and for the metallicity analysis in the next paper, we attempted a crossmatch between this catalogue and those of Savino and Sarajedini to search for any RRLs missing from ours. Of the 406 found by Savino, 388 were recovered in our own variable search. Of the remaining 18 stars, two were not found at all during data reduction, and the other 16 all failed to be recovered due to a variability index below 6. Applying our classification process to these stars, we determined that two were not true variables, while the remaining 14 are; these are listed with their relevant parameters from our own analysis in Table \ref{tab:savino_stars}, along with the Savino period for comparison. Unfortunately, a similar comparison with Sarajedini is not possible, as we were unable to crossmatch the reported coordinates with our own due to a systematic non-linear offset between the two coordinate systems.

\begin{figure}
    \centering
    \includegraphics[width=1\linewidth]{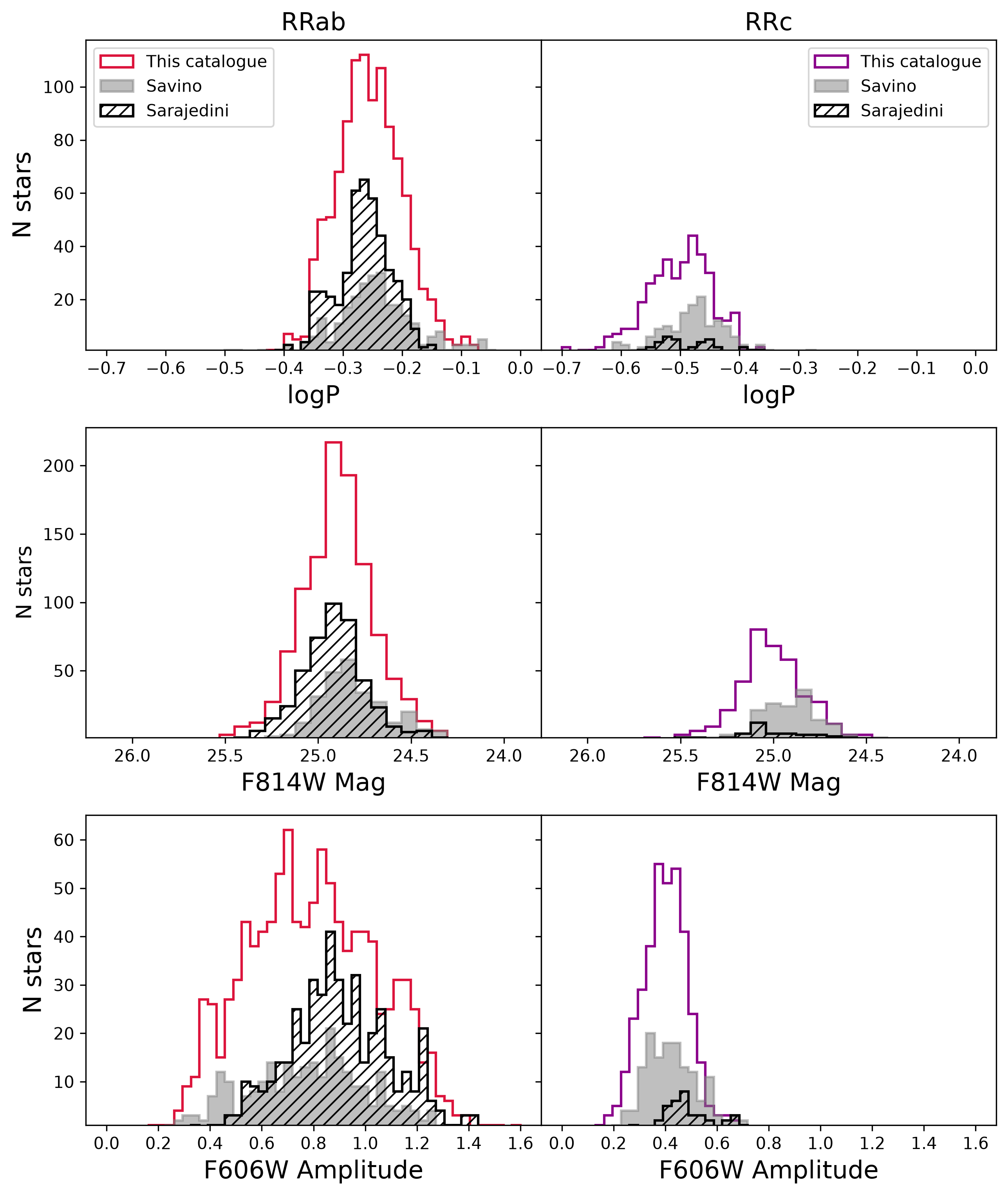}
    \caption{Distributions of period, F814W intensity-mean magnitude, and F606W amplitude for the RRab (left) and RRc (right) in this catalogue (coloured outlines), compared with those of \citet{savino-2022} (grey shaded) and \citet{sarajedini-2025} (hatched).}
    \label{fig:hist_compare}
\end{figure}

The different RRL populations across the three catalogues lead to slightly different mean parameters for the NGC 205 RRL, as shown in Fig.~\ref{fig:hist_compare} and summarised in Table \ref{tab:mean_compare}. For RRab, our mean period of 0.56 d and F606W amplitude of 0.80 mag both fall between the Savino and Sarajedini values, while for RRc our mean period of 0.32 d and mean amplitude of 0.40 mag are shorter than both the Savino and Sarajedini results. The mean magnitudes follow a different pattern: in both modes our values agree closely with Sarajedini but are around 0.1 mag fainter than Savino's. The similarity between our mean magnitudes and those of Sarajedini, contrasted with the larger discrepancy with Savino, warrants closer inspection. 

\begin{table}
\centering
\caption{Mean RR Lyrae parameters in NGC 205 from this work, \citet{savino-2022}, and \citet{sarajedini-2025}}
\label{tab:mean_compare}
\begin{tabular}{llccc}
\hline
\hline
Parameter & Mode & This work & Savino & Sarajedini \\
\hline
\multirow{2}{*}{$ P $ (d)} & RRab & 0.56 & 0.58 & 0.55 \\
 & RRc & 0.32 & 0.34 & 0.33 \\
\hline
\multirow{2}{*}{$ A_{606}$ (mag)} & RRab & 0.80 & 0.78 & 0.90 \\
 & RRc & 0.40 & 0.43 & 0.49 \\
\hline
\multirow{2}{*}{$ m_{814} $ (mag)} & RRab & 24.89 & 24.80 & 24.92 \\
 & RRc & 25.01 & 24.90 & 25.00 \\
\hline
\hline
\end{tabular}
\end{table}

Since the discrepancy in mean magnitudes could result from differing RRL samples, we performed a second comparison of magnitudes using only the stars in common between this catalogue and Savino's. For each matching star, we calculated the difference in intensity-mean magnitudes between the two catalogues. In F814W, the median of these differences was 0.066 mag with a normalised median absolute deviation (NMAD) of 0.069 mag, while in F606W, the median difference was 0.055 mag with an NMAD of 0.064 mag, confirming a true offset in the reported magnitudes. To determine whether this offset arises from different fitting methods or data reduction procedures, we also performed an epoch-by-epoch magnitude difference comparison for the same stars. This produced median differences of 0.050 and 0.047 mag with NMADs of 0.073 mag in both bands, indicating that the magnitude offset stems primarily from a difference in photometric zero-point rather than in the fitting procedure.

\begin{figure}
    \centering
    \includegraphics[width=1\linewidth]{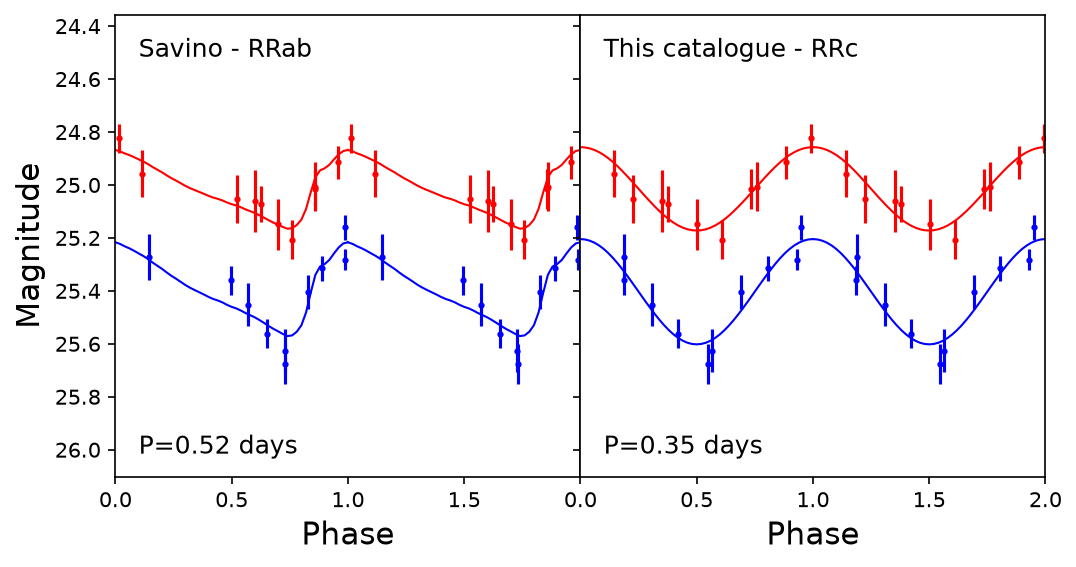}
    \caption{Phased light curves for star 203909 (233219 in the \citealt{savino-2022} catalogue), fitted at the Savino period of 0.52 d as an RRab (left) and at our period of 0.35 d as an RRc (right). Red and blue correspond to the F814W and F606W photometry, respectively.}
    \label{fig:lc_compare}
\end{figure}

Finally, among the stars in common with Savino's, roughly 25 per cent had disagreements in period, often changing the star's assigned pulsation mode. Frequently, neither fit is obviously preferred, either visually or by $\chi^2$. Fig.~\ref{fig:lc_compare} shows one such star. Since Savino's per-star template assignments are not publicly available, we performed an \textsc{rrfit} search over a small range around the Savino-reported period using our own templates. Both fits produce comparably good light curves, and each places the star squarely within the expected region of both the PL plot and Bailey diagram; its position near the centre of the horizontal branch is consistent with either mode, and so does not help distinguish between the two fits. In this case, and many others, it is difficult to determine which fit is correct. This is likely a result of the limited number of epochs and the cadence at which the data were taken; additional epochs would be required to break the degeneracy.

\section{Conclusions} \label{sec:conc}

We have presented the first paper in a series aimed at characterising the variable star population of NGC 205 based on \emph{HST} ACS imaging in the F606W and F814W filters. In this paper, we have described the determination of the distance to NGC 205, the identification and classification of variable stars by type and pulsation mode, the derivation of their pulsation properties, and the validation of these results. Our main results are as follows.  

We measure a tip of the RGB magnitude of $F814W({\rm TRGB}) = 20.64 \pm 0.01$, corresponding to a distance modulus of \(\mu = 24.68 \pm 0.06\), in agreement with previously published estimates.

We present a catalogue of 1441 RRL stars, 35 ACs, five RRL/AC stars, and 105 likely variables of uncertain type. This catalogue is substantially more complete than previous ones for this galaxy, which reported between four and five hundred RRL stars. Candidate variables were identified using the Stetson variability index, and preliminary pulsation periods, amplitudes, and intensity-mean magnitudes were derived with \textsc{rrfit}. Variables were classified by type using a combination of the light-curve fit, position in the CMD, position in the PL plane, and, for RRL, position in the Bailey diagram. For the confirmed variables, we ran an MCMC to sample the posterior, adopting the maximum-posterior parameters and taking uncertainties from the sampled distributions. 

Using a sample of 2000 simulated RRLs processed through the same pipeline, we find that 12 per cent fail the variability index cut, with low-amplitude stars the most likely to be missed. We further find that our alias-removal procedure substantially improves period and mode recovery, particularly for RRc. After this step, 96 per cent of recovered stars have periods within 5 per cent of their input values.

Of the RRLs, 1064 are fundamental mode (RRab) and 363 first-overtone (RRc), with the remainder either lacking a clear mode or admitting acceptable fits at two periods. They range in period from 0.200 to 0.836 d with mean periods of 0.561 d for RRab and 0.317 d for RRc. Roughly 25 per cent of the RRL are first-overtone pulsators, which, together with the position of the RRab in the Bailey diagram, indicates an Oosterhoff intermediate classification with a substantial metal-rich RRL population. This is reinforced by the 118 HASPs we identify (11 per cent of the RRab stars), since these high-amplitude short-period stars form only in sufficiently metal-rich environments.

Combining our RRL catalogue with a count of RGB stars, we measure $\log(N_\mathrm{RRL}/N_\mathrm{RGB}) = -1.24$. This places NGC 205 among galaxies with extended star formation histories, in agreement with the star formation history derived by \citet{savino-2025}. The published RRL catalogues of \citet{savino-2022} and \citet{sarajedini-2025} would instead give ratios near $-1.75$, which would imply a substantially smaller ancient population than the one we detect, illustrating the importance of catalogue completeness.

The 35 ACs place NGC 205 among the slow evolvers on the relation between AC frequency and host luminosity, indicating that both AC formation channels have contributed. This finding confirms the extended star formation history suggested by the fraction of RR Lyrae and HASP stars present in the galaxy.

Comparison with theoretical models in the CMD constrains their metallicity: the faintest ACs are reproduced by models at $[\mathrm{Fe/H}] = -2.1$ to $-2.6$, with poorer agreement at higher metallicity. The ACs are distributed uniformly across the field, indicating that metal-poor star formation was not confined to a single region.Although the old population (as traced by RRLs) enriched in metals relatively quickly on average, the spatial distribution of ACs highlights the inhomogeneous nature of this enrichment. Consequently, metal-poor gas clouds were still capable of forming stars at intermediate ages (i.e., ${\sim}$5 Gyr ago).

The catalogue presented here forms the basis for the analysis in the forthcoming companion paper where we will use the periods and intensity mean magnitudes of the RRLs to derive metallicities via a PLZ relation, and examine the implications for the chemical enrichment history of NGC 205.

\section*{Acknowledgements}

This research has made use of the NASA/IPAC Extragalactic Database, which is funded by the National Aeronautics and Space Administration and operated by the California Institute of Technology. We also acknowledge the use of NASA's Astrophysics Data System Bibliographic Services. 

This work is based on observations obtained from the Mikulski Archive for Space Telescopes (MAST). STScI is operated by the Association of Universities for Research in Astronomy, Inc., under NASA contract NAS5–26555. Support to MAST for these data is provided by the NASA Office of Space Science via grant NAG5–7584 and by other grants and contracts.

M.Mo., G.F., G.B. acknowledges support from the project ``ASTRA: Across Space and Time: Relics \& Archaeology'' (P.I. M. Marconi) funded by INAF-Instituto Nazionale di Astrofisica.
M.Mo. and G.F. acknowledge the INAF projects "Participation in LSST - Large Synoptic Survey Telescope" (LSST inkind contribution ITA-INA-S22, PI: G. Fiorentino), OB.FU. 1.05.03.06 and "MINI-GRANTS (2023) DI RSN2" (PI: G. Fiorentino), OB.FU. 1.05.23.04.02.
This research was supported by the International Space Science Institute (ISSI) in Bern/Beijing through ISSI/ISSI-BJ International Team project ID \#24-603 - "EXPANDING Universe" (EXploiting Precision AstroNomical Distance INdicators in the Gaia Universe).

I.P. gratefully acknowledges Grace Gephart for her help with data inspection and her discovery of several new variable star candidates.

%%%%%%%%%%%%%%%%%%%%%%%%%%%%%%%%%%%%%%%%%%%%%%%%%%
\section*{Data Availability}
All of the \emph{HST} data presented in this paper were obtained from the Mikulski Archive for Space Telescopes (MAST) at the Space Telescope Science Institute. The specific observations analysed can be accessed via \href{https://doi.org/10.17909/8dy0-6n20}{doi:10.17909/8dy0-6n20}. Galactic extinction values were taken from the NASA/IPAC Extragalactic Database and can be accessed through \href{https://doi.org/10.26132/NED1}{doi:10.26132/NED1}.

The \textsc{variable identification} and \textsc{variable classification} software used in this work are publicly available as part of Pulsating Variable Tools, version 1, which can be accessed through Zenodo at \href{https://doi.org/10.5281/zenodo.5152590}{doi:10.5281/zenodo.5152590}. The \textsc{rrfit} software is not publicly available but may be obtained from the authors of \citet{yang-2011} upon reasonable request.

Tables~\ref{tab:rrl_full} and \ref{tab:savino_stars} are published in their entirety in machine-readable form in the online supplementary material accompanying this article, and are also available from the Centre de Données astronomiques de Strasbourg (CDS) via VizieR.


\bibliographystyle{mnras}
\bibliography{paperI} 

%%%%%%%%%%%%%%%%%%%%%%%%%%%%%%%%%%%%%%%%%%%%%%%%%%

%%%%%%%%%%%%%%%%% APPENDICES %%%%%%%%%%%%%%%%%%%%%

\appendix
\section{The Catalogue} \label{app:catalogue}

\begin{landscape}

\begin{table}
\centering
\caption{Full properties of the variables in NGC 205. RA and Dec are given in degrees (J2000); P is the pulsation period in days; t$_{0}$ is the epoch of maximum light in MJD; m$_{606}$ and m$_{814}$ are intensity-mean magnitudes and A$_{606}$ and A$_{814}$ the corresponding amplitudes, all in mag; $\chi^{2}$ is the reduced $\chi^{2}$ of the best-fit solution; type is the variable class; and mode is the pulsation mode (FU for fundamental pulsators and FO for first-overtone pulsators). Subscripts $-$ and $+$ denote the lower and upper uncertainties on the preceding quantity. The flag column marks stars with two acceptable period solutions as 2F; both fits appear as separate entries. Stars of intermediate period where both RRab and RRc templates fit acceptably are marked UM. PF indicates poor fits or poor photometry, and UC indicates unclassified variables. This table is published in its entirety in machine-readable format; a portion is shown here for guidance regarding its form and content.}
\label{tab:rrl_full}
\setlength{\tabcolsep}{3pt}
\begin{tabular}{cccccccccccccccccccccccc}
\hline
\hline
ID & RA & Dec & P & P$_{-}$ & P$_{+}$ & t$_{0}$ & m$_{606}$ & m$_{606,-}$ & m$_{606,+}$ & A$_{606}$ & A$_{606,-}$ & A$_{606,+}$ & m$_{814}$ & m$_{814,-}$ & m$_{814,+}$ & A$_{814}$ & A$_{814,-}$ & A$_{814,+}$ & $\chi^{2}$ & Type & Mode & Flag \\
\hline
1283 & 10.03554 & 41.74185 & 0.32838 & 0.00476 & 0.00485 & 59135.32510 & 25.245 & 0.024 & 0.017 & 0.398 & 0.054 & 0.047 & 24.930 & 0.026 & 0.024 & 0.424 & 0.069 & 0.058 & 0.48 & RRL & FO &  \\
1503 & 10.03007 & 41.74956 & 0.67434 & 0.01327 & 0.00888 & 59134.97479 & 25.289 & 0.019 & 0.011 & 0.379 & 0.047 & 0.031 & 24.825 & 0.021 & 0.015 & 0.277 & 0.038 & 0.046 & 1.54 & RRL & FU &  \\
1894 & 10.02905 & 41.75110 & 0.55476 & 0.02286 & 0.03437 & 59135.35213 & 25.147 & 0.017 & 0.026 & 0.317 & 0.065 & 0.047 & 24.455 & 0.027 & 0.027 & 0.318 & 0.062 & 0.072 & 1.16 & -- & -- & UC \\
2136 & 10.03144 & 41.74760 & 0.55452 & 0.01431 & 0.01655 & 59135.34978 & 25.290 & 0.024 & 0.037 & 1.010 & 0.077 & 0.060 & 24.893 & 0.034 & 0.023 & 0.675 & 0.092 & 0.057 & 0.47 & RRL & FU &  \\
2307 & 10.02079 & 41.76254 & 0.31207 & 0.00418 & 0.00365 & 59135.30501 & 25.458 & 0.015 & 0.017 & 0.455 & 0.028 & 0.047 & 25.114 & 0.018 & 0.021 & 0.341 & 0.041 & 0.058 & 1.71 & RRL & FO &  \\
2390 & 10.02926 & 41.75107 & 0.66919 & 0.00885 & 0.01274 & 59135.06914 & 25.235 & 0.016 & 0.022 & 0.488 & 0.064 & 0.040 & 24.740 & 0.030 & 0.018 & 0.362 & 0.058 & 0.069 & 0.79 & RRL & FU &  \\
2651 & 10.02211 & 41.76091 & 0.56621 & 0.01506 & 0.01993 & 59135.31978 & 25.201 & 0.013 & 0.016 & 0.644 & 0.038 & 0.038 & 24.675 & 0.014 & 0.017 & 0.421 & 0.039 & 0.036 & 1.61 & RRL & FU &  \\
4891 & 10.02376 & 41.75980 & 0.54935 & 0.00745 & 0.02289 & 59135.30086 & 25.189 & 0.037 & 0.009 & 0.842 & 0.044 & 0.052 & 24.735 & 0.032 & 0.016 & 0.491 & 0.050 & 0.059 & 0.63 & RRL & FU &  \\
5156 & 10.02266 & 41.76143 & 0.28147 & 0.00342 & 0.00369 & 59135.43549 & 25.365 & 0.015 & 0.018 & 0.381 & 0.048 & 0.045 & 24.862 & 0.017 & 0.017 & 0.251 & 0.042 & 0.036 & 1.06 & RRL & FO &  \\
5246 & 10.03019 & 41.75124 & 0.62251 & 0.00935 & 0.01514 & 59135.13842 & 24.593 & 0.019 & 0.025 & 0.444 & 0.064 & 0.060 & 23.972 & 0.020 & 0.028 & 0.375 & 0.077 & 0.049 & 0.50 & AC & FU &  \\
8456 & 10.02882 & 41.75468 & 0.70060 & 0.03006 & 0.01972 & 59135.44442 & 25.306 & 0.018 & 0.017 & 0.338 & 0.037 & 0.056 & 24.776 & 0.023 & 0.016 & 0.183 & 0.061 & 0.039 & 0.85 & RRL & FU & PF \\
18166 & 10.03073 & 41.75689 & 0.29647 & 0.00476 & 0.00400 & 59135.23150 & 25.426 & 0.018 & 0.019 & 0.484 & 0.031 & 0.055 & 25.126 & 0.016 & 0.023 & 0.307 & 0.034 & 0.060 & 2.22 & RRL & FO & 2F \\
18166 & 10.03073 & 41.75689 & 0.42702 & 0.00684 & 0.00839 & 59135.45445 & 25.427 & 0.019 & 0.015 & 0.589 & 0.055 & 0.041 & 25.138 & 0.016 & 0.021 & 0.539 & 0.076 & 0.038 & 2.19 & RRL & FU & 2F \\
538704 & 10.05208 & 41.73775 & 0.38131 & 0.00173 & 0.00254 & 59135.23877 & 25.253 & 0.016 & 0.018 & 0.703 & 0.066 & 0.041 & 24.853 & 0.023 & 0.016 & 0.403 & 0.037 & 0.073 & 1.87 & RRL & -- & UM \\
\hline
\hline
\end{tabular}
\end{table}

\begin{table}
\centering
\caption{Properties of the RR Lyrae recovered from the Savino catalogue that were not identified in our own variable search. P$_{\rm S}$ is the period reported by Savino; all other quantities are from our analysis. RA and Dec are given in degrees (J2000); P is the pulsation period in days; t$_{0}$ is the epoch of maximum light in MJD; m$_{606}$ and m$_{814}$ are intensity-mean magnitudes and A$_{606}$ and A$_{814}$ the corresponding amplitudes, all in mag; $\chi^{2}$ is the reduced $\chi^{2}$ of the best-fit solution; type is the variable class; and mode is the pulsation mode (FU for fundamental pulsators and FO for first-overtone pulsators). Subscripts $-$ and $+$ denote the lower and upper uncertainties on the preceding quantity. A machine-readable version of this table is also available, and includes the corresponding Savino identifiers.}
\label{tab:savino_stars}
\setlength{\tabcolsep}{3pt}
\begin{tabular}{ccccccccccccccccccccccccc}
\hline
\hline
ID & RA & Dec & P$_{\rm S}$ & P & P$_{-}$ & P$_{+}$ & t$_{0}$ & m$_{606}$ & m$_{606,-}$ & m$_{606,+}$ & A$_{606}$ & A$_{606,-}$ & A$_{606,+}$ & m$_{814}$ & m$_{814,-}$ & m$_{814,+}$ & A$_{814}$ & A$_{814,-}$ & A$_{814,+}$ & $\chi^{2}$ & Type & Mode \\
\hline
16037 & 10.03707 & 41.74420 & 0.51826 & 0.51944 & 0.00712 & 0.02634 & 59135.07270 & 24.957 & 0.025 & 0.030 & 0.401 & 0.063 & 0.052 & 24.498 & 0.051 & 0.026 & 0.198 & 0.073 & 0.070 & 1.12 & RRL & FU \\
27573 & 10.04149 & 41.74171 & 0.38670 & 0.38556 & 0.00651 & 0.00798 & 59135.42365 & 25.396 & 0.023 & 0.027 & 0.268 & 0.066 & 0.050 & 24.955 & 0.030 & 0.023 & 0.333 & 0.072 & 0.049 & 0.86 & RRL & FO \\
49428 & 10.02993 & 41.76397 & 0.34298 & 0.51641 & 0.00420 & 0.01374 & 59135.24412 & 25.181 & 0.014 & 0.036 & 0.616 & 0.079 & 0.045 & 24.884 & 0.016 & 0.024 & 0.206 & 0.063 & 0.050 & 2.79 & RRL & FU \\
161882 & 10.06097 & 41.75576 & 0.40207 & 0.67463 & 0.02108 & 0.00937 & 59134.96351 & 25.342 & 0.020 & 0.023 & 0.239 & 0.051 & 0.061 & 24.932 & 0.026 & 0.024 & 0.296 & 0.065 & 0.061 & 1.51 & RRL & FU \\
333136 & 10.09319 & 41.76403 & 0.32423 & 0.32400 & 0.00354 & 0.00493 & 59135.21543 & 25.565 & 0.019 & 0.028 & 0.560 & 0.054 & 0.044 & 25.208 & 0.022 & 0.033 & 0.133 & 0.043 & 0.077 & 1.04 & RRL & FO \\
377963 & 10.09183 & 41.78052 & 0.59054 & 0.59808 & 0.00810 & 0.00897 & 59134.97865 & 25.625 & 0.039 & 0.033 & 0.743 & 0.080 & 0.068 & 25.169 & 0.046 & 0.041 & 0.512 & 0.077 & 0.075 & 2.93 & RRL & FU \\
381151 & 10.10272 & 41.76611 & 0.34556 & 0.25586 & 0.00467 & 0.00396 & 59135.40286 & 25.397 & 0.020 & 0.021 & 0.316 & 0.043 & 0.046 & 25.025 & 0.026 & 0.022 & 0.159 & 0.070 & 0.046 & 1.08 & RRL & FO \\
668799 & 10.07376 & 41.74239 & 0.25455 & 0.33994 & 0.00432 & 0.00605 & 59135.22904 & 25.505 & 0.024 & 0.024 & 0.409 & 0.058 & 0.057 & 25.193 & 0.027 & 0.028 & 0.270 & 0.068 & 0.053 & 0.80 & RRL & FO \\
679629 & 10.07253 & 41.74742 & 0.83766 & 0.61528 & 0.00947 & 0.01752 & 59134.94014 & 25.418 & 0.020 & 0.026 & 0.410 & 0.053 & 0.064 & 24.851 & 0.023 & 0.022 & 0.177 & 0.067 & 0.052 & 1.54 & RRL & FU \\
782027 & 10.09474 & 41.74859 & 0.62742 & 0.62826 & 0.01241 & 0.00514 & 59134.88850 & 25.222 & 0.020 & 0.015 & 0.592 & 0.043 & 0.064 & 24.786 & 0.029 & 0.019 & 0.245 & 0.081 & 0.050 & 1.11 & RRL & FU \\
784128 & 10.10452 & 41.73569 & 0.82148 & 0.38016 & 0.00575 & 0.00496 & 59135.14058 & 25.003 & 0.017 & 0.019 & 0.277 & 0.040 & 0.046 & 24.642 & 0.025 & 0.017 & 0.168 & 0.037 & 0.060 & 1.00 & RRL & FO \\
834116 & 10.11351 & 41.73931 & 0.47950 & 0.48008 & 0.00247 & 0.00232 & 59135.14163 & 25.605 & 0.017 & 0.026 & 0.863 & 0.059 & 0.055 & 25.236 & 0.033 & 0.028 & 0.562 & 0.067 & 0.071 & 2.60 & RRL & FU \\
851246 & 10.10948 & 41.75070 & 0.37426 & 0.37421 & 0.00365 & 0.00327 & 59135.19543 & 25.265 & 0.016 & 0.017 & 0.417 & 0.058 & 0.029 & 24.932 & 0.022 & 0.017 & 0.126 & 0.050 & 0.055 & 1.30 & RRL & FO \\
861701 & 10.11221 & 41.75042 & 0.63615 & 0.63114 & 0.01024 & 0.00889 & 59135.46034 & 25.424 & 0.016 & 0.024 & 0.584 & 0.061 & 0.075 & 24.945 & 0.017 & 0.021 & 0.201 & 0.075 & 0.056 & 1.81 & RRL & FU \\

\hline
\hline
\end{tabular}
\end{table}
\end{landscape}
%%%%%%%%%%%%%%%%%%%%%%%%%%%%%%%%%%%%%%%%%%%%%%%%%%

% Don't change these lines
\bsp	% typesetting comment
\label{lastpage}
\end{document}